\documentclass[12pt,a4paper]{article} % use larger type; default would be 10pt

\usepackage[utf8]{inputenc} % set input encoding (not needed with XeLaTeX)

\usepackage{amssymb}
\usepackage{mathrsfs}
\usepackage{bbm}
\usepackage{amsmath}

\usepackage{amssymb}
\usepackage{graphicx}%
\usepackage{dcolumn}%
\usepackage{bm}%
\usepackage{multirow}
\usepackage{afterpage}
\usepackage{float}

\usepackage{geometry} % to change the page dimensions
\usepackage{setspace}
\usepackage{fullpage}
\usepackage{graphicx} % support the \includegraphics command and options

\usepackage{authblk}
\usepackage{enumitem}
\usepackage{amsmath}
\usepackage{bm}
\usepackage{booktabs} % for much better looking tables
\usepackage{array} % for better arrays (eg matrices) in maths
\usepackage{paralist} % very flexible & customisable lists (eg. enumerate/itemize, etc.)
\usepackage{verbatim} % adds environment for commenting out blocks of text & for better verbatim======
\usepackage{subfig} % make it possible to include more than one captioned figure/table in a single float
\usepackage{fancyhdr} % This should be set AFTER setting up the page geometry
\usepackage{sectsty}
\allsectionsfont{\sffamily\mdseries\upshape} % (See the fntguide.pdf for font help)
\usepackage[nottoc,notlof,notlot]{tocbibind} % Put the bibliography in the ToC
\usepackage[titles,subfigure]{tocloft} % Alter the style of the Table of Contents

\begin{document}

\title{Model Risk Measurement under Wasserstein Distance}
\author[$\dagger$]{Yu Feng}
\author[$\dagger$]{Erik Schl\"ogl}

\affil[$\dagger$]{University of Technology Sydney, Quantitative Finance Research Centre}

\maketitle

\abstract{The paper proposes a new approach to model risk measurement based on the Wasserstein distance between two probability measures. It formulates the theoretical motivation resulting from the interpretation of fictitious adversary of robust risk management. The proposed approach accounts for equivalent and non-equivalent probability measures and incorporates the economic reality of the fictitious adversary. 
It provides practically feasible results that overcome the restriction %and the integrability issue imposed 
of considering only models implying probability measures equivalent to 
the reference model. The Wasserstein approach suits for various types of model risk problems, ranging from the single-asset hedging risk problem to the multi-asset allocation problem. The robust capital market line, accounting for the correlation risk, is not achievable with other non-parametric approaches.}
\section{Introduction}
Most current work on robust risk management either focuses on parameter uncertainty or relies on comparison between models. To go beyond that, Glasserman and Xu recently proposed a non-parametric approach \cite{glasserman2014robust}. Under this framework, a worst-case model is found among all alternative models in a neighborhood of the reference model. Glasserman and Xu adopted the Kullback-Leibler divergence (i.e. relative entropy) to measure the distance between an alternative model and the reference model. They also proposed the use of the $\alpha$-divergence to avoid heavy tails that causes integrability issues under the Kullback-Leibler divergence.

\begin{figure}[H]
\begin{center}
\includegraphics[width=.45\textwidth]{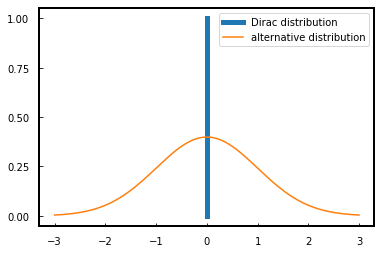}
\includegraphics[width=.45\textwidth]{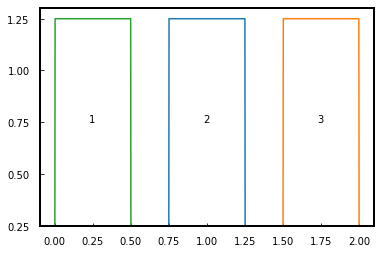}
\renewcommand{\figurename}{Fig}
\caption{(a) Dirac measure has a support of a single point. An alternative model with a widespread distribution cannot be related to the reference model using $f$-divergence. (b) State transition in a metric space. $f$-divergence does not involve the metric, so the transition from State 1 to 2 takes the same amount of cost as the transition from 1 to 3. 
%Remember to redraw the diagram to lift the base line from 0 to 0.2
}\label{intro}
\end{center}
\end{figure}

Both the Kullback-Leibler divergence and the $\alpha$-divergence are special examples of the $f$-divergence \cite{ali1966general, csisz1967information, ahmadi2012entropic}.  
A big problem of using $f$-divergence is that it is well-defined only when the alternative measure is absolutely continuous with respect to the reference measure. This limits the range of the alternative models under consideration. In some cases, we may want to search over all possible probability measures, whether they are absolutely continuous or not. This is especially true when we apply this approach to volatility, which corresponds to the quadratic variation of a process. If the process is driven by a Brownian motion, then searching over absolutely continuous measures rules out any model risk with respect to volatility. In Fig.~\ref{intro}(a), the distribution of the volatility is a Dirac-$\delta$ function under the reference model. The worst-case scenario that accounts for the volatility risk has a widely spread distribution of the volatility. However, $f$-divergence is not well-defined in this case, and therefore the worst-case scenario simply gets ignored.

Furthermore, the state space considered by financial practitioners is usually equipped with a natural metric. For instance, the price of a security takes value from the set of positive real numbers, and thus naturally inherits the Euclidean metric. Assuming a diffusion process, the price of the security moves along a continuous path. This means that a large price change is less probable than a small price change, implying a stronger deviation from the reference model. However, the distance of the move, measured by the natural metric, is not explicitly taken into account when using $f$-divergence. %In fact, by ignoring the metric of the state space, one would treat the two transitions indifferently. %When applying $f$-divergence to measure the model risk, we are not accounting for the transition cost from one state to another. This is counter-intuitive, as any transition takes a path.
Fig.~\ref{intro}(b) shows three models corresponding to three distributions of the security price. Assuming the Model 1 is adopted as the reference model, then Model 2 as an alternative model is apparently more probable than Model 3. However, one cannot tell the difference using any type of $f$-divergence, as the models have disjoint support.

As an attempt to solve these issues, we suggest to adopt the Wasserstein metric to measure the distance between probability measures. Relying on the metric equipped in the state space, the Wasserstein metric works for any two measures, even if their supports are mutually exclusive. As a result, the proposed Wassertein approach accounts for all alternative measures instead of merely the absolutely continuous ones. These features allow us to resolve the two issues of the $f$-divergence as mentioned above. For financial practitioners, the proposed approach is especially useful when dealing with reference measures with a subspace support (such as a Dirac measure).  %The introduction of transition cost keeps the change of support to a minimum level.

This paper is organized in the following manner. Sec.~\ref{sec:ad} offers a conceptual introduction including the intuitive motivation and the basics about the Wasserstein metric and its associated transportation theory. Sec.~\ref{sec:theory} is the theoretical part that provides the problem formulation and main results. It also includes practical considerations and comparison between different approaches. Sec.~\ref{sec:app} gives a few interesting applications in mathematical finance, ranging from the volatility risk in option pricing and hedging to robust portfolio optimisation.

\section{Basic Concepts}
\subsection{Motivation and Adversary Interpretation}
\label{sec:ad}
To illustrate the idea of model risk in an intuitive way, we start from a simple discrete-state space. An example is the credit rating which is ordinal, e.g. A+, A, A-, BBB+, etc. Assuming we have a reference model that states that in a month the credit rating of an institution could be A+, A- or BBB+. The reference model assigns probabilities of 25\%, 50\% and 25\% to the three states. Since we do not possess complete information, model risk exists either because the actual probabilities of the three states are different or because other ratings are still possible. Glasserman and Xu proposed the so-called ``adversary'' interpretation which suggests a fictitious adversary that perturbs the probabilities against us \cite{glasserman2014robust}. By perturbing the probabilities essentially the adversary adds new information, limited by its information entropy budget. For example, if the adversary would like to move 5\% chance from A+ to BBB+, its consumption of relative entropy is
\begin{align}
0.2\ln\left(\frac{0.2}{0.25}\right)+0.3\ln\left(\frac{0.3}{0.25}\right)=0.01
\end{align}
Now suppose the adversary would like to move the 5\% chance to BBB, which is a state of 0 probability under the reference measure. The consumption of relative entropy 
\begin{align}
0.2\ln\left(\frac{0.2}{0.25}\right)+0.05\ln\left(\frac{0.05}{0}\right)
\end{align}
becomes infinite. This simply means that such perturbation is impossible no matter how much control the adversary has. %Another issue is on moving probability out of the two states. Even shifting 1\% chance from BBB+ to BBB is not allowed under Glasserman and Xu's approach.
In the language of probability theory, relative entropy is well-defined only when the new measure is absolutely continuous with respect to the nominal one.

To allow for a more generic quantification of model risk, we may re-define the requested cost of perturbation. Instead of using the relative entropy, we consider about the cost of a state transition (termed as the transportation cost). This transportation cost is usually given by some metric on the state space. For simplicity we assume that the distance between two credit ratings is given by the number of ratings in between, e.g. $d($A+, A-$)$=2 and $d($A+, BBB+$)$=3. %downgrading the rating by 1 level always incurs 1 unit of cost to the adversary. By doing this we essentially provide a metric to the state space. 
We calculate the weighted average transportation costs for the two types of perturbations discussed in the last paragraph:\\
1. shift 5\% chance from A+ to BBB+: transportation cost=5\%$\times$3=0.15\\
%2. shift 25\% chance from A+ to BBB+: transportation cost=25\%$\times$2=0.5\\
2. shift 5\% chance from A+ to BBB: transportation cost=5\%$\times$4=0.2\\
The second-type perturbation only involves a cost slighter larger than the first type, instead of being infinite.

Using the transportation cost described above, one can measure the adversary's cost for all alternative measures rather than merely the absolutely continuous ones. It may provide state transitions that are highly concentrated. %, thus not reflecting the actual competitive market structure. 
To illustrate this point, think about the transition from state A+. The fictitious adversary would push the rating only in one direction. This implies that the transportation performed by the fictitious agent can be represented by a (deterministic) map on the state space $T:\Omega\to\Omega$. $T$ is called a \textbf{transportation map} \cite{Villani1}. In fact, suppose it is optimal for the fictitious agent (thus the worst case scenario) to transit the state A+ to a state, say BBB+. There is no motivation for the agent to transport any probability mass from A+ to other states. This results from the linearity of the transportation cost, and will be illustrated further in the next section. %As a result, the model risk.

Glasserman and Xu's interpretation of model risk involves an fictitious adversary but without explicit consideration of its economic nature. They assume that the adversary performs uniformly aiming to maximise our expected loss. In reality, such an adversary can only be achieved by a single agent or institution. The actual market structure, however, is usually more competitive. In economic terms, the fictitious adversary may consist of heterogeneous agents who act independently.  This asks for approaches that quantify the model risk based on the actual market structure. 

Now get back to the credit rating example. In reality there might be multiple agents that are capable of impacting the rating, among which some prefer to upgrade the rating while others prefer to dowgrade the rating. This asks for a different formulation of state transitions, for the final state transited from a given initial state becomes a random variable. All we know is a probability measure conditional to the given initial state (or a transition density). Overall, the transportation is described by a joint probability density $\gamma:\Omega\times\Omega\to\mathbb{R}^+$ instead of a deterministic map. The joint density (or the corresponding measure on $\Omega\times\Omega$) is refered as the \textbf{transportation plan}\cite{Villani2}. This allows us to formulate the optimisation problem w.r.t the transportation plan instead of the transportation map. Such formulation leads to more general results capable of accounting for different types of market structure.% that takes the optimisation w.r.t transportation map as its special case. This point will be illustrated later after the main result is derived.

From a practical perspective, the main advantage of using the Wasserstein metric is to deal with reference measures supported by strict subspaces. Still in the example of credit rating, the reference measure is supported by $\{A+,A-,BBB+\}$, which is a strict subspace of the entire state space (of rating). Approaches based on $f$-divergence are only capable of incorporating alternative measures with the same support. Using the Wasserstein approach, on the other hand, does allow us to alter the support. In particular, if we formulate the problem using a transportation map $T$, then the new support is $\{T(A+),T(A-),T(BBB+)\}$, still a strict subspace. Therefore, although different transportation maps provide us with different supports, none of them is capable of spreading to the entire state space. On the other hand, by formulating the problem with a transportation plan, we indeed account for alternative measures that are supported by the entire space. Now regarding the fictitious adversary as a class of hetergenuous agents, it is reasonable to believe that the distribution is widely spread under the perturbation of the adversary. %For example, the transition density from state $A+$ is uniform for all ratings above $A+$. The transition density from $BBB+$ is uniform for all ratings below $BBB+$. The support of the new measure thus covers the entire state space. Assuming hetergemuous agents reduces the concentration of final states, resulting in a widely spread alternative probability measure. 

Thus, we are interested in an approach to model risk measurement that formulates the transportation cost based on a transportation plan. We will see that this approach is capable of account for actual market structure by parametrising an entropy constraint (Sec.~\ref{sec:entropy}). In the remaining part of this section, we will review the Wasserstein metric and its associated transportation theory. 

\subsection{Transportation Theory and Wasserstein Metric}\label{sec:was}

Starting from this point, we will always assume a continuous-state space unless otherwise stated. The approach for discrete-state spaces follows the same routine and therefore is omitted. 
Now let the state space ($\Omega$, $d$) be a Polish metric space, we may define the transportation cost $c:\Omega\times\Omega\to\mathbb{R}_+$ by the $n$-th power of the metric, i.e. $c(x,y)=d(x,y)^n$, where $n\in[1,\infty)$. 
Given two probability measures $P$ and $Q$ on ($\Omega$, $d$), we may formulate the optimal transportation problem using either a transportation map or a transportation plan. For the former approach, we aim to find the transportation map $T:\Omega\to\Omega$ that realizes the infimum
\begin{align}\label{eq:monge}
\inf_T &\int_\Omega p(x)c\left(x,T(x)\right)dx\\
s.t.&~\left|J_T(x)\right|q\left(T(x)\right)=p(x),~\forall x\in\Omega\nonumber
\end{align}
where $p(x)$ and $q(x)$ are the probability density functions of the two measures $P$ and $Q$, respectively. 
%Usually $d$ is given by a metric assigned to the state space $\Omega$. When $\Omega$ is an Euclidean space, $d$ may take the form of the corresponding Euclidean metric (or some monotonic function of the Euclidean metric).
$J_T$ is the Jacobian of the map $T$. It is part of the constraint that enforces the map $T$ to be measure-preserving. 
%The metric function $c(x,y)$ also defines the transportation cost from a state $x$ to another state $y$. 
Eq.~\ref{eq:monge} is refered as the \textbf{Monge's formulation} of the optimal transportation problem. 

The problem of Monge's formulation is that the existence of a measure-preserving map $T$ is not guaranteed. Examples in the last section provide a discrete-state illustration of this issue: $supp(Q)=\{T(A+),T(A-),T(BBB+)\}$ has at most three elements. As a result, there is no measure-preserving map if $|supp(Q)|>|supp(P)|$. %the support of $Q$ must share the same cardinality.
In a continuous-state space, a measure-preserving map sends a Dirac measure to another Dirac measure. Therefore, measure-preserving map does not exist if $P$ is a Dirac measure while $Q$ is not. %We may improve t
The ill-posed Monge's formulation can be improved by adopting a transportation plan $\gamma:\Omega\times\Omega\to\mathbb{R}_+$:
\begin{align}\label{eq:w1}
\inf_{\gamma}&\int_{\Omega\times\Omega}\gamma(x,y)c(x,y)dxdy\\
s.t.&\int_{\Omega}\gamma(x,y)dy=p(x)\nonumber\\
&\int_{\Omega}\gamma(x,y)dx=q(y)\nonumber
\end{align}
Eq.~\ref{eq:w1} is refered as the \textbf{Kantorovich's formulation} of the optimal transportation problem. 
It is clear that every transportation map $T$ can be given by a transportation plan
\begin{align}\label{eq:relation}
\gamma(x,y)=\left|J_T(x)\right|q\left(y\right)\delta\left(y-T(x)\right)
\end{align}
where $\delta(\cdot)$ is the Dirac-$\delta$ function. In addition, the existence of a transportation plan is guaranteed as $\gamma(x,y)=p(x)q(y)$ always satisfies the constraints in Eq.~\ref{eq:w1}. According to these observations, the Kantorovich's formulation is preferred over  the Monge's formulation. 
Remember that the transportation cost $c(x,y)$ is the $n$-th power of the metric $c(x,y)$. The $n$-th \textbf{Wasserstein matric}, denoted by $W_n$, is defined as the infimum in Eq.~\ref{eq:w1}, raised to the power of $1/n$. In the next section, the theoretical formulation and the main results of this paper will be presented with the help of the Kantorovich's formulation. The transportation cost function $c(x,y)$ will be regarded as a generic non-negative function, without reference to its specific form or the power $n$.

\section{Theory}

\subsection{Wasserstein Formulation of the Model Risk Problem}\label{sec:theory}
The core part of model risk measurement is to determine the alternative model under the worst-case scenario. In the language of probability theory, we need to determine the alternative probability measure that maximizes our expected loss. 
We may formulate the problem in the following way. %using the concepts discussed in the previous paragraphs. 
Given a nominal probability measure $P$ on the state space $\Omega$, we would like to find a worst-case measure $Q^*$ that realizes the following supremum: 
\begin{align}\label{eq:basic}
\sup_{Q}\,&\mathsf{E}^Q[V(X)]\\
s.t.\,&D(P||Q)\leq\eta\nonumber
\end{align}
The expectation is taken under the alternative measure $Q$, on a loss function $V:\Omega\to\mathbb{R}$. 
%The state variable $X$ is a random variable that takes value from $\Omega$. 
Only alternative measures that are close enough to the reference measure are deemed as legitimate. This restriction is formulated by constraining the statistical distance $D(P||Q)$ to be equal to or less than a constant $\eta$. 

Glasserman and Xu suggest using the relative entropy (or Kullback-Leibler divergence) for $D(P,Q)$. Like any $f$-divergence, relative entropy has limited feasibility as only equivalent measures are legitimate. Based on the discussion in the last section, we suggest to apply the Wasserstein metric instead. 
The actual formulation of the model risk problem, on the other hand, has a slightly different form than Eq.~\ref{eq:basic}. Specifically, instead of optimizing the expectation w.r.t the alternative measure $Q$ (or its density function $q:\Omega\to\mathbb{R}^+$), we optimize the expectation w.r.t the transportation plan $\gamma:\Omega\times\Omega\to\mathbb{R}^+$ directly. The single constraint on $q$ is replaced by two constraints applied to $\gamma$, including the marginalisation condition given in Eq.~\ref{eq:w1}. This formulation is based on the idea of state transition and is illustrated below.
%The concept of Wasserstein metric deeply roots from the optimal transportation theory. 

Based on the discussion in the last section, for any pair of states $x,y\in\Omega$ all we need to find is the transition density from $x$ to $y$, $p_{Y|X}(y|x)$. Given a function of transportation cost from $x$ to $y$, $c(x,y)$, the expected transportation cost conditional to an initial state $x$ is %given by
\begin{align}
W(x)=\int_\Omega p_{Y|X}(y|x)c(x,y)dy
\end{align}
The initial state $x$ follows a distribution $p_X(x)$ given by the reference model. Take expectation under the reference measure, we get the unconditional transportation cost
\begin{align}\label{eq:distances}
W=\int_\Omega p_X(x)W(x)dx=\int_{\Omega\times\Omega}p_{X,Y}(x,y)c(x,y)dxdy
\end{align}
where the joint distribution $p_{X,Y}(x,y)=p_X(x)p_{Y|X}(y|x)$. To be consistent with the notation used previously, we denote the marginal distributions $p_X$, $p_Y$ by $p$, $q$, and the joint distribution $p_{X,Y}$ by the transportation plan $\gamma$. It is noted that the transition converts the initial distribution $p(x)$ to a final distribution $q(y)$, inducing a change of measure on the state space $\Omega$. 

One of the key tasks of the model risk measurement is to solve for the worst-case model under certain constraints. These constraints set the criteria for legitimate  alternative models. Now denote the loss function by $V(x)$ ($x\in\Omega$), the probability density function of the reference model by $p(x)$, and the probability density function of an alternative model by $q(x)$. 
We formulate the problem by the supremum of the expected loss over all legitimate models:
\begin{align}\label{eq:prob0}
\sup_{q(y)}\int_{\Omega} q(y)V(y)dy
\end{align}
According to the discussion in the last section, we regard the change of measure as probabilistic state transitions. %, with the transition probability denoted by the conditional probabily $\gamma(y|x)$. 
The probability density function $q(y)$ of the alternative model is merely the marginalisation of a joint density (or transportation plan) $\gamma(x,y)$, i.e. $
q(y)=\int_{\Omega} \gamma(x,y)dx$. 
This allows us to take the supremum over $\gamma(x,y)$ instead of $q(y)$:
\begin{align}\label{eq:prob}
\sup_{\gamma(x,y)}\int_{\Omega\times\Omega} \gamma(x,y)V(y)dxdy
\end{align}The first constraint of the supremum problem comes from the marginalisation of the joint density w.r.t $x$, as it is given by the reference model:
\begin{align}\label{eq:constraints1}
\int_{\Omega} \gamma(x,y)dy=p(x)
\end{align}
In a similar way to Glasserman and Xu's work, we restrict all alternative measures by their distances from the reference model. The distance is now measured by the average transportation cost given in Eq.~\ref{eq:distances}. It reflects the expected cost paid by a fictitious adversary who attempts to transit a state $x$ to an alternative state $y$ according to the transportation plan $\gamma(x,y)$. This results in the following constraint which defines the set of legitimate measures:
\begin{align}\label{eq:constraints2}
\int_{\Omega\times\Omega} \gamma(x,y)c(x,y)dxdy\leq& \eta
%\int_{\Omega\times\Omega} q(x,y)\ln q(x,y)dxdy\leq& \mu\nonumber
\end{align}
 %$q(x,y)$ is a bivariate probability density function on the product space $\Omega\times\Omega$. $q(y)=E_x[q(x,y)]$ is the probability density function of an alternative measure on $\Omega$.
%$c(x,y)$ measures the cost for a fictitious adversary to transit a state $x$ to an alternative state $y$. Usually $d$ is given by a metric assigned to the state space $\Omega$. When $\Omega$ is an Euclidean space, $d$ may take the form of the corresponding Euclidean metric (or some monotonic function of the Euclidean metric).

%The intepretation of the three constraints will be provided in the following paragraphs.
%Among the three constraints (Eq.~\ref{eq:constraints}), the first two 
%The two constraints combined gives 
The constant $\eta$ in Eq.~\ref{eq:constraints2} is termed as the Wasserstein distance budget, just as the relative entropy budget in Glasserman and Xu's approach. In order to account for a specific density function $q^*(y)$ in the constrained supremum problem given by Eq.~\ref{eq:prob0}-\ref{eq:constraints2}, the Wasserstein distance, defined in Eq.~\ref{eq:w1}, between $q^*(y)$ and the nominal density $p(x)$ cannot exceed $\eta$.  %i.e. legitimate alternative measures should have a Wasserstein distance equal to or less than $\eta$ to the reference measure. 
In fact, if $q^*(y)$ can be obtained by marginalizing a transportation cost $\gamma^*(x,y)$ that satisfies Eq.~\ref{eq:constraints1}-\ref{eq:constraints2}, then according to Eq.~\ref{eq:w1} its Wasserstein distance with the nominal density function $p(x)$ is
\begin{align}
W(p,q^*)=&\inf_{\gamma}\int_{\Omega\times\Omega}\gamma(x,y)c(x,y)dxdy\nonumber\\
\leq&\int_{\Omega\times\Omega} \gamma^*(x,y)c(x,y)dxdy\leq \eta
\end{align}
% does not satisfy the first two constraints. 
%To illustrate the point, consider the Wasserstein distance $W_1$ defined in Eq.~\ref{eq:w1}. %between two probability measures on $\Omega$, with density functions denoted by $p(x)$ and $q(y)$. It is defined by% \cite{}
%The relation between the Wasserstein metric and the constraints formulated in Eq.~\ref{eq:constraints1}-\ref{eq:constraints2} is illustrated as below. 
On the other hand, if $W(p,q^*)<\eta$, then the density function $q^*(y)$ can always be expressed by %$p'(x)=\int_{\Omega}q(x,y)dx$ny alternative measure with probability density function of $q$ can be expressed by 
the marginalisation of a transportation plan $\gamma^*(x,y)$ that satisfies Eq.~\ref{eq:constraints1}-\ref{eq:constraints2}. Otherwise, in the definition of the Wasserstein distance, Eq.~\ref{eq:w1}, $\eta$ sets a lower bound for the term
\begin{align}\label{eq:constraints20}
\int_{\Omega\times\Omega}\gamma(x,y)c(x,y)dxdy
\end{align}
Therefore the Wasserstein distance, as the infimum of the term above, is equal to or larger than $\eta$. This immediately violates the assumption $W(p,q^*)<\eta$. 
%$\int_{\Omega\times\Omega} q(x,y)c(x,y)dxdy\leq\eta$. Such alternative measure should satisfy the first two constraints. 
In summary, $\eta$ sets the maximum level (budget) of Wasserstein distance for an alternative measure to be legitimate.

Remarkably, even though the problem (Eq.~\ref{eq:prob}-\ref{eq:constraints2}) is formulated using the transportation plan (Kantorovich's formulation), its solution can be expressed by a transportation map $T^*:\Omega\to\Omega$,
\begin{align}\label{eq:ystar}
T^*(x)=\arg\max_{y\in\Omega}\left[V(y)-\frac{c(x,y)}{\beta}\right]
\end{align}
where $\beta\in\mathbb{R}_{++}$ is a constant. The underlying reason is the linearity of Eq.~\ref{eq:constraints20} w.r.t the transportation plan $\gamma$. Suppose the worst case scenario is to transit a state $x$ to another state $T^*(x)$. Then there is no motivation for the fictitious adversary to transit $x$ to states other than $T^*(x)$, say $T'(x)$, for the adversary could continue improving the target by increasing $\gamma(x,T^*(x))$ while reducing $\gamma(x,T'(x))$ (by the same amount). See Appendix A for a sketch of the derivation of Eq.~\ref{eq:ystar}. 

\subsection{Entropy Constraint on Transportation Plan}
\label{sec:entropy}
Eq.~\ref{eq:ystar} provides the worst-case transportation map for the problem formulated in Eq.~\ref{eq:prob}-\ref{eq:constraints2}. This formulation in fact assumes a zero-sum game between two parties, in which our counterparty attempts to shift a state $x$ to $y^*(x)$ (deterministically) so that its profit (thus our loss) can be maximized. In Sec.~\ref{sec:ad}, we mentioned that the actual market structure may be more competitive consisting of heterogeous agents that act more or less independently. This calls for a widespread transition density $p_{Y|X}(y|x)$ (instead of being a $\delta$-function).

%In the last section, we have mentioned about the advantage of assuming multiple agents in the adversary interpreation. This also has practical benefits. 

In practice, it is also advantageous of having a widely distributed transition density. 
For the purpose of risk management, we need to consider a wide range of alternative measures due to model ambiguity. As a result a widespread distribution is usually more representative than a narrow distribution. From the information-theoretic point of view, a widespread distribution contains less information (more entropy) thus more appropriately representing the model ambiguity.  
Now have a think about the practical situations where the approaches based on $f$-divergence are not applicable. They usually have reference measures that are too restrictive in the sense that they are supported by merely subspaces (of the state space). 
To correctly quantify the model risk one should consider widespread distributions supported by the entire state space. However, these distributions do not have well-defined $f$-divergence w.r.t the reference measure, providing an inherent issue of these approaches.% measurement using $f$-divergence is incomplete, for it simply excludes all measures that are not equivalent (including those 

One of the primary purposes of using Wasserstein metric instead of $f$-divergence is to tackle this issue. Specifically, we would like to include all measures regardless of their support. This purpose is achieved by using the Kantorovich's formulation as illustrated in Sec.~\ref{sec:was}.
%the transportation plan, but not the transportation map. In fact, transportation map restricts the space of alternative measures, for sometimes there exists no transportation map that links two measures (see Sec.~\ref{sec:was}). This is the reasonable why we formulate the problem using the transportation plan.
However, without further constraint the worst-case model can still be achieved with a transportation map, as illustrated by Eq.~\ref{eq:ystar}. This causes the worst-case measure to be restrictive if the reference measure is supported by merely a subspace. To achieve a widespread worst-case distribution, one may need to impose further constraints to Eq.~\ref{eq:prob}-\ref{eq:constraints2}. 

%In some practical applications, a worst-case model induced by a transportation map $T$ might not be desirable. 
A Dirac reference measure, denoted by $P$, provides a special example where Eq.~\ref{eq:ystar} is not suitable for characterizing the worst-case scenario. Applying the transportation map $T^*$ results in the worst-case measure supported by $\{T(x)\}$ where $x$ is the sole element in $supp(P)$. The worst-case measure is Dirac as well. 
In most cases, this worst-case measure inappropriately accounts for model ambiguity. 
%is given by the support of the reference measure $P$ via a map $T$. This, however, does not guarantee that $supp(Q)$ covers the entire state space.
%When we investigate the potential model risks, we don't want to be that restrictive. 
To resolve this issue, we may further impose an entropy constraint that guarantees the worst-case measure to be supported by the entire state space: %This requirement is not satisfied using $f$-divergences as they are not capable of altering the support. It might not be met by the Wasserstein distance budget as well. Intuitively, assuming the adversary is just one agent. It would transit a state $x$ to another state $y$ deterministically, if such transition incurs the least amount of transportation cost. There is no motivation to transit $x$ partly to $y$ and partly to $z$ due to the linearity of the transportation co
%This requirement may be fulfilled by imposing an entropy constraint:
%Introducing the condition of heterogeneous agents resolves the issue. This calls for a third constraint:
\begin{align}\label{eq:constraints3}
-\int_{\Omega\times\Omega} \gamma(x,y)\ln \gamma(x,y)dxdy\geq \mu
\end{align}
The LHS is the (differential) entropy \cite{cover2012elements} of the joint distribution (transportation plan) $\gamma(x,y)$, and the RHS is a constant $\mu\in\mathbb{R}$ (or a positive constant $\mu\in\mathbb{R}_{++}$ for discrete-state space). This constraint excludes every transportation plan that is equivalent to a transportation map. In fact, every transportation map $T$ gives a transportation plan with a $\delta$-function transition density (see Eq.~\ref{eq:relation}). For such transportation plan, the $\delta$-function makes the LHS of Eq.~\ref{eq:constraints3} 
approaching negative infinity (or zero for discrete-state space), and is therefore excluded.

%Formally, this constraint restricts the entropy of the joint probability density function. If the LHS is too large, the alternative density function cannot cover the entire state space, and is therefore insufficient in terms of model ambiguity (violating the purpose of model risk management). 
Alternatively, Eq.~\ref{eq:constraints3} can be interpreted with respect to the transition density function $p_{Y|X}(y|x)$. We may rewrite Eq.~\ref{eq:constraints3} by
\begin{align}\label{eq:constraints31}
-\int_{\Omega\times\Omega} \gamma(x,y)\ln p_{Y|X}(y|x)dxdy\geq \mu+\int_{\Omega} p(x)\ln p(x)dx
\end{align} 
Eq.~\ref{eq:constraints31} imposes a restriction on the transition density function. 
A tighter restriction (with a larger $\mu$) implies a wider transition density, reflecting a market structure that is more competitive. On the other hand, if we relax the constraint completely by shifting $\mu$ towards negative infinity (or zero for discrete-state space), then we permit transition densities to take the form of $\delta$-functions, corresponding to the single-agent adversary. 

We may further introduce terms from information theory, and rewrite Eq.~\ref{eq:constraints31} by
\begin{align}\label{eq:constraints32}
%\mu\geq&\int_{\Omega\times\Omega} q(x,y)d\ln q(x,y)dxdy\nonumber\\=&
%\int_{\Omega\times\Omega} q(x,y)d\ln \frac{q(x,y)}{p(x)q(y)}dxdy+\int_{\Omega} p(x)\ln p(x)dx+\int_{\Omega} q(y)\ln q(y)dy\nonumber\\
%=&I(X,Y)-H(X)-H(Y)
%-\int_{\Omega\times\Omega}\gamma(x,y)\left[\ln p(x)+\ln p_{Y|X}(y|x)\right]dxdy=
\int_\Omega p(x)H(Y|X=x)dx\geq\mu-H(X)
\end{align}
%where $p_{Y|X}(y|x)$ %=\gamma(x,y)/p(x)$ 
%is the conditional probability density (transition density from $x$ to $y$). 
%We can see that the LHS of Eq.~\ref{eq:constraints3} contains two parts. The first part is the entropy w.r.t $p(x)$, 
where $H(X)$ denotes the entropy of the random variable $X$ \cite{cover2012elements}. Since its distribution $p(x)$ is given by the reference model, $H(X)$ is deemed as a constant. $H(Y|X=x)$, on the other hand, is the information entropy w.r.t the transition density $p_{Y|X}(y|x)$. It is interpreted as the entropy of the random variable $Y$, conditional to $X$ taking a given value $x$. $H(Y|X=x)$ quantifies the uncertainty of the transportation from a given state $x$. Generally a more competitive market that involves more independent decision-makers leads to a more uncertain state transition, thus a larger $H(Y|X=x)$. 
As a result, Eq.~\ref{eq:constraints32} allows us to incorporate the actual market structure by parametrising $\mu$. 
%Since we have a family of such density functions as $x$ can take any value from $\Omega$, we need to aggregate their corresponding entropies. This is done by taking their expected value (weighted by $p(x)$). Since $H(X)$ is a constant, the constraint Eq.~\ref{eq:constraints3} is equivalent to a lower bound on the expected entropy (or uncertainty) of transitions:
%If theses density functions $p_{Y|X}(y|x)$ have shapes that are too narrow, their corresponding entropies $H(Y|X=x)$ get large, thus violating Eq.~\ref{eq:constraints4}.
 %$H(Y|X=x)$ is the entropy of the random variable Y conditional to $X$ taking a certain value. 
It is noted that in information theory, the LHS of Eq.~\ref{eq:constraints4} is termed as the conditional (differential) entropy and is denoted by $H(Y|X)$ \cite{cover2012elements}. 
This leads to an equivalent information-theoretic version of the constraint Eq.~\ref{eq:constraints3}:
\begin{align}\label{eq:constraints4}
H(Y|X)\geq\mu-H(X)
\end{align}

\subsection{Main Result and Discussion}

The supremum problem Eq.~\ref{eq:prob}, subject to the three constraints Eq.~\ref{eq:constraints1}, \ref{eq:constraints2} and \ref{eq:constraints3}, formulates the complete version of the Wasserstein approach to model risk measurement. 
Now suppose there exists a joint distribution $\gamma^*(x,y)$ that solves the problem. Then the worst-case model is characterised by a probability density function 
\begin{align}\label{eq:qstar}
q^*(y)=\int_{x\in\Omega}\gamma^*(x,y)dx,~~\forall y\in\Omega
\end{align}
%The model risk (in terms of the expected loss under worst-case alternative measure) is calculated accordingly by $\int_{x\in\Omega}q^*(y)V(y)dy$.
To solve the constrained supremum problem, we introduce two multipliers $\alpha\in\mathbb{R}_+$ and $\beta\in\mathbb{R}_+$, and transform the original problem to a dual problem. Solving the inner part of the dual problem leads to our main result (see Appendix B for derivation):
\begin{align}\label{eq:main}
q^*(y)=&\int_\Omega dx\frac{p(x)\exp\left(\frac{V(y)}{\alpha}-\frac{c(x,y)}{\alpha\beta }\right)}{\int_\Omega\exp\left(\frac{V(z)}{\alpha}-\frac{c(x,z)}{\alpha\beta }\right)dz}%\right|X=x\rig\mathbb{E}_X\left(\frac{\exp\left(\frac{V(y)}{\alpha}-\frac{c(x,y)}{\alpha\beta }\right)}{\mathbb{E}_Y\left[\left.\exp\left(\frac{V(y)}{\alpha}-\frac{c(x,y)}{\alpha\beta }\right)\right|X=x\right]}\right)
\end{align}
%which the density function of the worst-case measure. 
It is noted that the multipliers $\alpha$ and $\beta$ are in fact controlling variables that determine the levels of restriction, of the entropy constraint Eq.~\ref{eq:constraints3} and the transportation constraint Eq.~\ref{eq:constraints2}, respectively.

The limit when $\alpha$ approaches zero corresponds to complete relaxation of the entropy constraint Eq.~\ref{eq:constraints3}. In this limit Eq.~\ref{eq:qstar} degenerates to the probability density function induced by the transportation map given by Eq.~\ref{eq:ystar}. On the other side of the spectrum, Eq.~\ref{eq:qstar} approaches a uniform distribution when $\alpha$ approaches infinity, as a result of the tight entropy constraint.

In the extreme case of $\beta=0$, Eq.~\ref{eq:qstar} leads to a simple result $q^*(x)=p(x)$. This is because the transportation constraint Eq.~\ref{eq:constraints2} reaches its tightest limit ($\eta=0$). No state transition is allowed thus preserving the reference model. On the other hand, when $\beta$ approaches infinity, the worst-case distribution $q^*(y)\sim \exp(V(y)/\alpha)$ is exponentially distributed. In this case, the transportation cost is essentially zero. As a result, the worst-case measure is the one that maximises the expected value of $V(Y)$ with a reasonably large entropy (the maximum expected value is given by a Dirac measure at $\arg\max_y V(y)$ but this results in a very low entropy). Special cases of Eq.~\ref{eq:main} are tabulated in Tab.~\ref{tab:1} for different values of $\alpha$ and $\beta$.

\begin{table}[H]
\centering
\caption{Worst-case probability density function at different $(\alpha, \beta)$ combinations. $p$ is the nominal distribution and $u$ is the uniform distribution. $\delta$ denotes the Dirac $\delta$-function and $T^*$ is the transportation map given by Eq.~\ref{eq:ystar}.}\label{tab:1}
\begin{tabular}{|m{2cm}||m{20ex}|m{20ex}|m{10ex}|}
		\hline
		 & $\alpha=0$ & $\alpha$ & $\alpha\to\infty$ \\[5pt]
		\hline
		\hline
		$\beta=0$&\multicolumn{3}{c|}{$p(x)$}\\[5pt]
		\hline
		$\beta$&$p(T^{^*-1}(x))/|J_T|$ & given by Eq.~\ref{eq:qstar}&\multirow{2}{*}{$\to u(x)$}\\[5pt]
		\cline{1-3}
		$\beta\to\infty$&$\delta(x-\arg\max V(x))$&$\propto e^{V(x)/\alpha}$&\\[5pt]
		\hline
	\end{tabular}
\end{table}

\subsection{Practical Considerations}

According to Table.~\ref{tab:1}, the worst-case measure approaches a uniform distribution when $\alpha$ approaches infinity (i.e. under the most restrictive entropy constraint). In practice, we may want the worst-case distribution to  converge to a given density function $q_0$ instead of being uniform. This requires modification on the formulation of the problem, by generalising the entropy constraint Eq.~\ref{eq:constraints4} to
\begin{align}\label{eq:constraints50}
-D_{KL}\left(P(Y|X)||Q_0(Y)\right)\geq\mu-H(X)-H(Y)
%I(X,Y)-H(p(x))+D_{KL}(q(Y)||q_0(Y))\leq\mu-\mathbb{E}(\ln q_0(Y))
\end{align}
$D_{KL}\left(P(Y|X)||Q_0(Y)\right)$ denotes the conditional relative entropy, given by the expected value of the KL divergence, $D_{KL}(P(Y|X=x)||Q_0(Y))$, of the two probability density functions w.r.t $y$, $p_{Y|X}(\cdot|x)$ and $q_0(\cdot)$. Written explicitly, the conditional relative entropy takes the form of
\begin{align}\label{eq:extra}
D_{KL}\left(P(Y|X)||Q_0(Y)\right)=&\int_{\Omega} p(x)\left(\int_{\Omega}p_{Y|X}(y|x)\ln\left(\frac{p_{Y|X}(y|x)}{q_0(y)}\right)dy\right)dx\nonumber\\
=&\int_{\Omega\times\Omega}\gamma(x,y)\ln\frac{\gamma(x,y)}{q_0(y)}dxdy-\int_\Omega p(x)\ln p(x)dx
\end{align}
Substituting Eq.~\ref{eq:extra} into Eq.~\ref{eq:constraints50} allows us to obtain the explicit version of the constraint:
\begin{align}\label{eq:constraints5}
-\int_{\Omega\times\Omega}\gamma(x,y)\ln\frac{\gamma(x,y)}{q_0(y)}dxdy-\int_\Omega q_0(y)\ln q_0(y)dy\geq\mu
\end{align}
It is clear that the previous entropy constraint Eq.~\ref{eq:constraints3} is merely a special case of Eq.~\ref{eq:constraints5} in which $q_0$ is a uniform distribution. %a relative entropy constraint. In fact, the LHS of Eq.~\ref{eq:constraints4} may be expressed in terms of the KL divergence $D(\cdot||\cdot)$:% it is equivalent to where the prior $q_0(y)$ is a uniform distribution. 
%We may therefore generalise the problem formulation in the last section by replacing the uniform distribution by any prior $q_0(y)$. 
Under this formulation, the problem that we need to solve consists of Eq.~\ref{eq:prob}, \ref{eq:constraints1}, \ref{eq:constraints2} and \ref{eq:constraints5}. The result differs from Eq.~\ref{eq:main} by a weighting function $q_0$ (see Appendix B for derivation):
\begin{align}\label{eq:8}
q^*(y)=&\int_\Omega dx\frac{p(x)q_0(y)\exp\left(\frac{V(y)}{\alpha}-\frac{c(x,y)}{\alpha\beta }\right)}{\int_\Omega q_0(z)\exp\left(\frac{V(z)}{\alpha}-\frac{c(x,z)}{\alpha\beta }\right)dz}
%q^*(y)=&\int_\Omega\left(\frac{q_0(y)\exp\left(\frac{V(y)}{\alpha}-\frac{c(x,y)}{\alpha\beta }\right)}{\mathbb{E}_Y\left[\left.\exp\left(\frac{V(y)}{\alpha}-\frac{c(x,y)}{\alpha\beta }\right)\right|X=x\right]}\right)
\end{align}

It is noted that Eq.~\ref{eq:8} takes a similar form to the Bayes' theorem and $q_0$ serves as the prior distribution. In fact, if the conditional distribution takes the following form:
\begin{align}\label{eq:bayes0}
p_{X|Y}^*(x|y)\propto \exp\left(\frac{V(y)}{\alpha}-\frac{c(x,y)}{\alpha\beta }\right)
\end{align}
Then the Bayes' theorem states that
\begin{align}\label{eq:bayes}
p_{Y|X}^*(y|x)=&\frac{p_{X|Y}^*(x|y)q_0(y)}{\mathsf{E}_Y\left(p_{X|Y}^*(x|\cdot)q_0(\cdot)\right)}\nonumber\\
=&\frac{q_0(y)\exp\left(\frac{V(y)}{\alpha}-\frac{c(x,y)}{\alpha\beta }\right)}{\int_\Omega q_0(z)\exp\left(\frac{V(z)}{\alpha}-\frac{c(x,z)}{\alpha\beta }\right)dz}
\end{align}
which is the posterior distribution of $Y$ given the observation $X=x$. Now if we observe a distribution $p(x)$ over $X$, then we may infer the distribution of $Y$ to be
\begin{align}\label{eq:bayes1}
q^*(y)=&\int_\Omega p(x)p_{Y|X}^*(y|x)dx\nonumber\\
=&\int_\Omega dx\frac{p(x)q_0(z)\exp\left(\frac{V(y)}{\alpha}-\frac{c(x,y)}{\alpha\beta }\right)}{\int_\Omega q_0(y)\exp\left(\frac{V(z)}{\alpha}-\frac{c(x,z)}{\alpha\beta }\right)dz}
\end{align}
which is exactly the worst-case distribution given in Eq.~\ref{eq:8}. 

The connection between the Bayes' theorem and Eq.~\ref{eq:8} is not just a coincidence.
In fact, the worst-case distribution of $Y$, given in Eq.~\ref{eq:bayes1}, can be regarded as the posterior distribution of a latent variable. On the other hand, the reference model of $X$, given by $p(x)$, is considered as the distribution that is actually observed. 
Assuming no reference model exists (i.e. no observation on $X$ has been made), then our best guess on the latent variable $Y$ is given solely by its prior distribution $q_0(y)$. Now if the observable variable $X$ does take a particular value $x$, then we need to update our estimation according to the Bayes' theorem (Eq.~\ref{eq:bayes}). The conditional probability density $p_{X|Y}^*(x|y)$ takes the form of Eq.~\ref{eq:bayes0}, reflecting the fact that the observable variable $X$ and the latent variable $Y$ are not far apart. Imagining that we generate a sampling set $\{x_i\}$ following the nominal distribution $p(x)$, then for each $x_i$ we get a posterior distribution $p_{Y|X}^*(y|x_i)$ from Eq.~\ref{eq:bayes}. Overall, the best estimation of the distribution over the latent variable $Y$ results from the aggregation of these posterior distributions. This is achieved by averaging them weighted by their probabilities $p(x_i)$, as given in Eq.~\ref{eq:bayes1}. This leads to the Bayesian interpretation of the model risk measurement, which concludes that by ``observing" the reference model $p(x)$ over the observable variable $X$, the worst-case model is given by updating the distribution of the latent variable $Y$, from the prior distribution $q_0(y)$ to the posterior distribution $q^*(y)$.

\begin{table}[!h]
\centering
\caption{Worst-case density function with prior $q_0$ at different $(\alpha, \beta)$ combinations. $p$ is the nominal distribution. $\delta$ denotes the Dirac $\delta$-function and $T^*$ is the transportation map given by Eq.~\ref{eq:ystar}.}\label{tab:2}
\begin{tabular}{|m{2cm}||m{20ex}|m{20ex}|m{10ex}|}
		\hline
		 & $\alpha=0$ & $\alpha$ & $\alpha\to\infty$ \\[5pt]
		\hline
		\hline
		$\beta=0$&\multicolumn{3}{c|}{$p(x)$}\\[5pt]
		\hline
		$\beta$&$p(T^{^*-1}(x))/|J_T|$ & given by Eq.~\ref{eq:8}&\multirow{2}{*}{$\to q_0(x)$}\\[5pt]
		\cline{1-3}
		$\beta\to\infty$&$\delta(x-\arg\max V(x))$&$\propto q_0(x)e^{V(x)/\alpha}$&\\[5pt]
		\hline
	\end{tabular}
\end{table}

If we know nothing about the reference model, setting the prior $q_0$ to a uniform distribution seems to make the most sense (because a uniform distribution maximizes the entropy thus containing least information). This leads to the main result given by Eq.~\ref{eq:main}. %and Tab.~\ref{tab:1}. 
However, it is sometimes much more convenient to choose a prior other than the uniform distribution. 
A particular interesting case is to set $q_0$ the same as the nominal distribution $p$. In this case, the limit of $\beta\to\infty$ (complete relaxation of the transportation constraint) is given by
\begin{align}\label{eq:speciallimit}
q^*(x)=\frac{p(x)e^{\theta V(x)}}{\int_\Omega p(x)e^{\theta V(x)}dx}
\end{align}
where we replace the parameter $\alpha^{-1}$ by $\theta$. This limit is exactly the worst-case distribution given by the relative entropy approach \cite{glasserman2014robust}. Despite of the simplicity of Eq.~\ref{eq:speciallimit}, %consistency with the relative entropy approach, 
it is not recommended to set $q_0=p$ because by doing so we lose the capability of altering the support of the reference measure. %This is the major issue of the relative entropy approach that we would like to resolve.

In practice, a common problem of the relative entropy approach is that the denominator in Eq.~\ref{eq:main} may not be integrable. %when $V(x)$ increases faster 
To see this point, we examine the worst-case density function under the relative entropy approach:
\begin{align}\label{eq:9}
q^*_{KL}(x)\propto
p(x)e^{\theta V(x)}
\end{align}
The RHS of Eq.~\ref{eq:9} may not be integrable if $V(x)$ increases too fast (or $p(x)$ decays too slowly as in the cases of heavy tails). As an example, we consider the worst-case variance problem where $V(x)=x^2$. If the reference model follows an exponential distribution, then Eq.~\ref{eq:9} is not integrable.

Using the proposed Wasserstein approach, however, the flexibility of choosing a proper prior $q_0$ helps us to bypass this issue. In fact, one may choose a prior distribution $q_0$, different from the nominal distribution $p$, to guarantee that it decays sufficiently fast. 
According to Eq.~\ref{eq:bayes}, all we need to guarantee is that
\begin{align}\label{eq:90}
q_0(y)\exp\left(\frac{V(y)}{\alpha}-\frac{c(x,y)}{\alpha\beta }\right)
\end{align}
is integrable w.r.t $y$. 
Fortunately, it is always possible to find some $q_0$ that satisfies this criteria. 
As a simple choice, we may set $q_0(y)\propto e^{-V(y)/\alpha}$ to ensure the integrability. Such choice makes Eq.~\ref{eq:90} proportional to
\begin{align}\label{eq:int}
%\int_\Omega 
\exp\left(-\frac{c(x,y)}{\alpha\beta}\right)
\end{align}
We suppose that the state space $\Omega$ is an Euclidean space with finite dimension and the transportation cost $c(x,y)$ is given by its Euclidean distance. Then for all $x\in\Omega$ Eq.~\ref{eq:int} is integrable w.r.t $y$, for the integrand diminishes exponentially when $y$ moves away from $x$.
% as $e^{-c(x,y)}p(x)$ converges faster than $p(x)$ and therefore is integrable. In the example above, we choose the alternative model prior $q_0(y)\propto e^{-\theta x^2}$ following a normal distribution, even if the reference model $p_0(x)$ follows an exponential distribution. The difference between the two distributions is penalised by the transportation cost carried by the metric $c(x,y)$.

In summary, formulating the problem using the relative entropy constraint Eq.~\ref{eq:constraints5} allows for flexibility of choosing a prior distribution $q_0$. This is practically useful as one can avoid integrability issue by selecting a proper prior. This flexibility is not shared by the relative entropy approach as in Glasserman and Xu \cite{glasserman2014robust}, which is regarded as a special case where the prior $q_0$ equals the nominal distribution $p$.
%Under the KL divergence, we know that heavy tail distribution could cause integrability issue. Under the Wasserstein approach, however, integrability can be guaranteed. 

%Eq.~\ref{eq:8} can be considered as a generalisation of the result under the KL divergene (Eq.~\ref{eq:9}). In fact, if we set $\beta\to \infty$ and $q_0(x)=p(x)$ for all $x\in\Omega$, then Eq.~\ref{eq:8} reduces to Eq.~\ref{eq:9} where $\theta=\alpha^{-1}$. By setting $\beta\to\infty$, we lose all the information about the metric (transportation cost from $x$ to $y$). By setting $q_0(x)=p(x)$ we lose the capability of altering the support of the original measure.

\section{Application}\label{sec:app}
\subsection{Jump risk under a diffusive reference model}
We start from a price process that takes the form of a geometric Brownian motion
\begin{align}
dS_t=\mu S_tdt+\sigma S_tdW_t
\end{align}
The logarithmic return at time $T$ follows a normal distribution:
\begin{align}
x:=\ln \left(\frac{S_T}{S_0}\right)\sim \mathcal{N}\left(\left(\mu-\frac{\sigma^2}{2}\right)T,\sigma^2T\right)
\end{align}
When the volatility reaches zero, the return becomes deterministic and the distribution density is
\begin{align}\label{eq:limit}
p(x)=\lim_{\sigma\to0}\frac{1}{\sqrt{2\pi T}\sigma}e^{-\frac{[x-(\mu-\sigma^2/2)T]^2}{2\sigma^2T}}=\delta(x-\mu T)
\end{align}
In the case, model risk cannot be quantified using $f$-divergence. In fact, the reference measure is a Dirac measure therefore no equivalent alternative measure exists. Under the KL divergence in particular, the worst-case measure is calculated by
\begin{align}
\frac{p(x)e^{\theta V(x)}}{\int_\Omega p(x)e^{\theta V(x)}dx}=\delta(x-\mu T)
\end{align}
which is the same as the reference measure. This is consistent with the Girsanov theorem for diffusion processes which states that the drift term is altered by some amount proportional to the volatility, i.e. $\tilde\mu=\mu-\lambda\sigma$. When the volatility under the reference model decreases to zero, the alternative measure becomes identical to the reference measure.

Approaches based on $f$-divergence excludes the existence of model risk given a zero volatility. This is, however, not true in practice, as the nominal diffusion process may still ``regime-switch'' to some discontinuous process. In fact, to quantify risks, one usually take into account the possibility of discontinuous changes of state variables (i.e. ``jumps''). Using the Wasserstein approach, quantifying such jump risk becomes possible, even if the reference model is based on a pure diffusion process. Substituting Eq.~\ref{eq:limit} into Eq.~\ref{eq:main} gives the worst-case distribution (see Appendix C for details)
\begin{align}\label{eq:jump}
q_W(x)=
\frac{\exp\left(\frac{V(x)}{\alpha}-\frac{c(x,\mu T)}{\alpha\beta }\right)}{\int_\Omega\exp\left(\frac{V(y)}{\alpha}-\frac{c(y,\mu T)}{\alpha\beta }\right)dy}
\end{align}

Notice that Eq.~\ref{eq:jump} is suitable for any application where the reference model is given by a Dirac measure. Under $f$-divergence, the limitation to equivalent measures keeps the reference model unchanged. The Wasserstein approach, on the other hand, relaxes such limitation, allowing for a worst-case model that differs from a Dirac measure. This allows us to measure risk in variables assumed to be deterministic in the reference model. A particularly interesting example is the quadratic variation process, which is deemed as deterministic under the Black-Scholes model. We will discuss this in detail later with regard to the model risk in dynamic hedging.

To illustrate Eq.~\ref{eq:jump}, we consider the expected value of $x$ under the worst-case scenario. This problem is formulated using Eq.~\ref{eq:basic} with a linear loss function $V(x)=x$. We further assume a quadratic transportation cost function $c(x,y)=(x-y)^2$. The worst-case distribution given by Eq.~\ref{eq:jump} turns out to be
\begin{align}\label{eq:linearex}
q_W(x)=\frac{1}{\sqrt{\pi\alpha\beta}}e^{-\frac{\left(x-\mu T-\beta/2\right)^2}{\alpha\beta}}
\end{align}
One can see that the worst-case scenario is associated with a constant shift of the mean (by $-\beta/2$), even if the reference measure is deterministic (i.e. Dirac). The change in mean is also associated with a proportional variance (i.e. $\alpha\beta/2$), if $\alpha$ is assigned a positive value. The resulting normal distribution, with a finite variance, is a reflection of model ambiguity. This is in contrast with approaches based on $f$-divergences, which are incapable of altering the reference model in this case, as its support includes only a single point.

\subsection{Volatility Risk and Variance Risk}

In this section, we consider the risk of volatility uncertainty given the nominal Black-Scholes model. When an option approaches maturity, the reference measure (on the price of its underlying asset) becomes close to a Dirac measure. This is visualised by the normal distribution of return narrowing in a rate of $\sqrt{t}$. When the time to maturity $t\to0$, the normal distribution shifts to a Dirac distribution with zero variance. %, so the alternative measure caused by jump can only be considered 

Under the Kullback-Leibler divergence (or any $f$-divergence), any model risk vanishes when the reference model converges to a Dirac measure. As a result, on a short time to maturity a sufficient amount of variance uncertainty can only be produced with a large cost (parametrised by $\theta$). To illustrate this point, consider a normal distribution (say Eq.~\ref{eq:limit} before taking the limit).
For the purpose of measuring the variance risk, we need to adopt a quadratic loss function $V(x)=x^2$. Under the Kullback-Leibler divergence, the variance of the worst-case distribution is given by \cite{glasserman2014robust}
\begin{align}\label{eq:klvariance}
\sigma_{KL}^2T=\frac{\sigma^2T}{1-2\theta\sigma^2T}
%(x)\propto\exp\left(-\frac{\left(y-\frac{\mu}{1-2\theta\sigma^2T}\right)^2}{\frac{2\sigma^2}{1-2\theta\sigma^2}}\right)
\end{align}
When time to maturity $T\to0$, the worst-case volatility $\sigma_{KL}\to\sigma$ with a fixed $\theta$. 
This is not consistent with what we see in the market. In fact, with short time to maturity the fear of jumps can play an important role. Such fear of risks is priced into options and variance swaps termed as the volatility (or variance) risk premium.

\begin{figure}[H]
\begin{center}
\includegraphics[width=.6\textwidth]{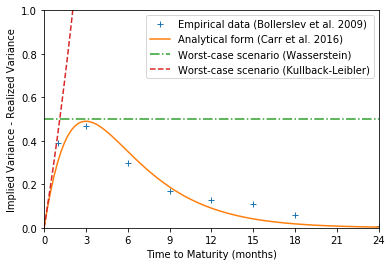}
\renewcommand{\figurename}{Fig}
\caption{Worst-case volatility as a function of time under the (a) Wasserstein approach, (b) KL divergence.}\label{fig:empirical}
\end{center}
\end{figure}

The volatility (or variance) risk premium can be considered as the compensation paid to option sellers for bearing the volatility risks \cite{bakshi2003delta,low2005volatility}. It is practically quantified as the difference between the implied volatility (or variance) and the realised volatilty (or variance). As it is priced based on the volatility risk, its quantity is directly linked to the risk associated with the reference measure used to model the underlying asset. Therefore by analyzing the term structure of such premium, one can get some insight into the worst-case volatility risk. Under the assumption of diffusive price dynamics, Carr and Wu developed a formula for the at-the-money implied variance \cite{carr2016analyzing}. Illustrated in Fig.~\ref{fig:empirical}, the formula matches well with the empirical data \cite{bollerslev2009expected} for maturities longer than 3 months. For maturities shorter than 3 months, however, the formula seems to underestimate the variance risk premium. Other empirical work also shows that option buyers consistently pay higher risk premium for shorter maturity options \cite{low2005volatility}.

The underestimation of volatility risk premium on short maturity is an intrinsic problem with diffusive models. Indeed, the work mentioned above reveals the importance of quantifying jumps when time to maturity remains short. Other work shows that the risk premium due to jumps is fairly constant across different maturities \cite{ait2015term}. This implies a very different time dependency from that due to continuous price moves (Eq.~\ref{eq:klvariance}). In fact, any approach based on $f$-divergence is incapable of producing sufficient model risk on $t\to0$, suggesting a decaying term structure of risk premium. On the other hand, the Wasserstein approach does not suffer from this issue. In fact, it produces a worst-case volatility that has little time dependence (Fig.~\ref{fig:empirical}). Therefore, the Wasserstein approach provides a particularly useful tool for managing the variance risk and quantifying its risk premium on short time to maturity. %, as the Euclidean metric defined on the real axis could not explode. We may 

\begin{comment}
To see this point, assume that the nominal distribution takes the form of 
\begin{align}
\frac{1}{\sqrt{2\pi}\sigma}e^{-\frac{1}{2}\left(\frac{x-x_0}{\sigma}\right)^2}
\end{align}
where the standard deviation $\sigma$ is very small. The worst-case distribution, under the relative entropy approach, is given by
\begin{align}
\frac{1}{\sqrt{2\pi}\sigma}e^{\theta V(x)-\frac{1}{2}\left(\frac{x-x_0}{\sigma}\right)^2}
\end{align}
The worst-case distribution converges to the nominal distribution on $\sigma\to0$ (i.e. $\delta$ function), in the sense that
\begin{align}
\lim_{\sigma\to0}E^Q(f(x))=&\lim_{\sigma\to0}\int_{-\infty}^\infty f(\sigma z+x_0)
\frac{1}{\sqrt{2\pi}}e^{\theta V(\sigma z+x_0)-\frac{1}{2}z^2}dz\\
=&f(x_0)
\end{align}
for arbitrary function $f:\mathbb{R}\to\mathbb{R}$. Under the Wasserstein approach, however, the worst-case measure is no longer a Dirac measure on $\sigma\to0$, showing consideration of jump risks. 
In fact, if we apply the quadratic type loss $V(x)=x^2$, the worst-case distribution according to Eq.~\ref{eq:jump} is
\begin{align}
q_W(y)\propto\exp\left(-\frac{\left(y-\frac{\mu}{1-\beta}\right)^2}{\frac{2\sigma^2}{(1-\beta)^2}+\frac{\alpha\beta}{(1-\beta)}}\right)
\end{align}
The volatility premium in percentage is
\begin{align}
\frac{\tilde\sigma}{\sigma}=\frac{1}{1-\beta}
\end{align}
\end{comment}

With the Wasserstein approach, the worst-case variance takes the form of (see Appendix B)
\begin{align}\label{eq:swt}
\sigma_W^2T=\frac{\sigma^2T}{(1-\beta)^2}+\frac{\alpha\beta}{2(1-\beta)}
\end{align}
The Wasserstein approach provides a worst-case variance that is independent of the time to maturity. It scales the nominal variance by a constant factor $(1-\beta)^{-2}$. In addition, it introduces a constant extra variance $\alpha\beta/(1-\beta)$. The extra variance term is modulated by the parameter $\alpha$. If we set $\alpha$ to zero, then the worst-case volatility $\sigma_W$ is merely a constant amplification of the nominal volatility $\sigma$. 
%This is consistent with the theoretical result derived for variance swap. 
This model risk measure, however, may not be sufficient if the nominal volatility is very close to zero. The extra variance term serves to account for the extra risks (e.g. jumps) that are not captured by the nominal volatility.

\begin{comment}
The worst-case variance may explain the volatility premium (the surplus of implied volatility over the realised volatility). Under the KL divergence, the volatility premium (as a percentage surplus) is given by
\begin{align}
\frac{\sigma_{KL}-\sigma}{\sigma}=\sqrt{\frac{1}{1-2\theta\sigma^2(T-t)}}-1
\end{align}
which decays to zero when approaching maturity. This is no longer a problem if switching to the Wasserstein approach. In fact, the worse-case variance under the Wasserstein approach contains two parts under which the volatility premium is a constant (assuming $\alpha=0$):
\begin{align}
\frac{\sigma_W}{\sigma}=\frac{\beta}{1-\beta}
\end{align}
With jump risk the volatility premium may even increase when approaching maturity. This is consistent with market observations.

\begin{figure}[H]
\begin{center}
\includegraphics[width=.45\textwidth]{wvol.png}
\includegraphics[width=.45\textwidth]{klvol.png}
\renewcommand{\figurename}{Fig}
\caption{Worst-case volatility as a function of time under the (a) Wasserstein approach, (b) KL divergence.}\label{fig:1}
\end{center}
\end{figure}

\begin{figure}[H]
\begin{center}
\includegraphics[width=.45\textwidth]{volpercent.png}
\includegraphics[width=.45\textwidth]{expirydist.png}
\renewcommand{\figurename}{Fig}
\caption{(a) Percentage volatility premium as the surplus of worst-case volatility over the nominal volatility and (b) Worst-case distributions under the two approaches.}\label{fig:2}
\end{center}
\end{figure}

\end{comment}

\subsection{Model Risk in Portfolio Variance}
The Wasserstein approach can be applied to quantify the risk associated with modelling the variance of a portfolio, assuming the asset returns follow a multivariate normal distribution. Suppose there are $n$ assets under consideration and their returns are reflected by a state vector $x$: $x\in\mathcal{V}$ where $\mathcal{V}$ is a n-dimensional vector space. For generality, we consider the following target function $V:\mathcal{V}\to\mathbb{R}_+$
\begin{align}\label{eq:xAx}
V(x)=x^TAx
\end{align}
where $A$ is a positive-definite symmetric matrix. If we replace $x$ by $x'=x-\mathsf{E}(x)$ and $A$ by $ww^T$, then the expected value of the target function reflects the portfolio variance:
\begin{align}
\mathsf{E}[V(x)]=\mathsf{E}(x^Tww^Tx)=w^T\Sigma w
\end{align}
where $w$ is the vector of compositions in the portfolio. $\Sigma$ is the covariance matrix of the normally distributed asset returns (under the reference model). 

\begin{figure}[H]
\begin{center}
\includegraphics[width=.3\textwidth]{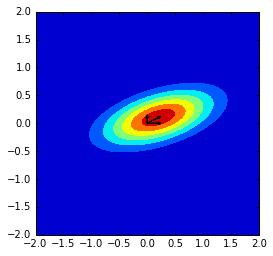}
\includegraphics[width=.3\textwidth]{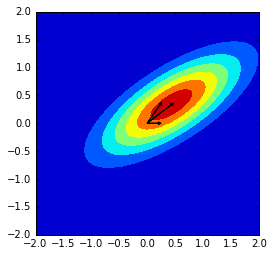}
\includegraphics[width=.3\textwidth]{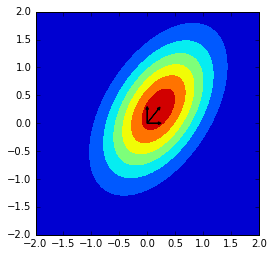}
\renewcommand{\figurename}{Fig}
\caption{Multivariate nominal distributions (a) reference model, (b) worst case under the KL divergence, (c) worst case under the Wasserstein approach (as a measure-preserving transform).}\label{fig:3}
\end{center}
\end{figure}

To find the worst-case model using the Wasserstein approach, we need to first define a metric in the vector space $\mathcal{V}$. Suppose the vector space is equipped by a norm $||x||$ then the metric is naturally defined by $c(x,y)=||x-y||$. Here we focus on the kind of norm that has an inner-product structure:
\begin{align}\label{eq:innerpro}
||x||=\sqrt{x^TBx}, ~\forall x\in\mathcal{V}
\end{align}
where $B$ is a positive-definite symmetric matrix (constant metric tensor). The resulting worst-case distribution is still multivariate normal, with the vector of means and covariance matrix replaced by (see Appendix D for derivation)
\begin{align}\label{eq:varw}
\mu_W=&(B-\beta A)^{-1}B\mu\\
\Sigma_W=&(B-\beta A)^{-1}B \Sigma B(B-\beta A)^{-1}+\frac{\alpha\beta}{2}(B-\beta A)^{-1}
\label{eq:sigmaw}
\end{align}
Apart from a constant term that vanishes if assigning zero to the parameter $\alpha$, the worst-case distribution is transformed from the nominal distribution via a measure-preserving linear map (see Appendix D). This result is more intuitive than the result obtained using the KL divergence, given by \cite{glasserman2014robust}
\begin{align}\label{eq:varkl}
\mu_{KL}=&(I-2\theta\Sigma A)^{-1}\mu\\
\Sigma_{KL}=&(I-2\theta\Sigma A)^{-1}\Sigma\label{eq:sigmakl}
\end{align}
Fig.~\ref{fig:3} provides an example illustrating that the worst-case distribution is indeed a measure-preserving transform with the Wasserstein approach.

\begin{figure}[t]
\begin{center}
\includegraphics[width=.3\textwidth]{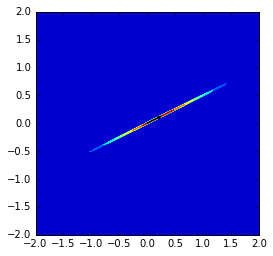}
\includegraphics[width=.3\textwidth]{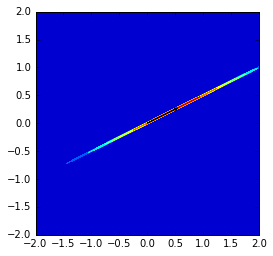}\\
\includegraphics[width=.3\textwidth]{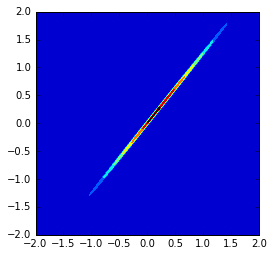}
\includegraphics[width=.3\textwidth]{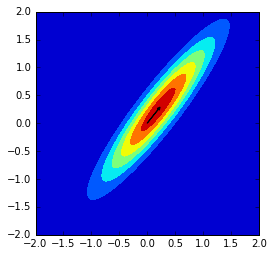}
\renewcommand{\figurename}{Fig}
\caption{Multivariate nominal distributions (a) reference model, (b) worst case under the KL divergence, when the support is a low-dimensional subspace. Worst-case multivariate nominal distributions under the Wasserstein approach (c) $\theta=0$ (d) $\theta=0.5$.}\label{fig:5}
\end{center}
\end{figure}

The constant term reflects residual uncertainty when the reference model has vanishing variances. This term is especially useful when some of the assets are perfectly correlated (either 1 or -1) and the vector space $\mathcal{V}$ is not fully supported by the reference measure. In this case, the Wasserstein approach provides results that differ significantly from the $f$-divergence approach. In particular, approaches based on KL divergence (or any f-divergences) cannot alter the support, they merely reweight the states within the support. This is illustrated in Fig.~\ref{fig:5}, where two assets are perfectly correlated. The reference model shown in (a) provides a measure supported by a one-dimensional vector subspace of $\mathcal{V}$. The worst-case measure under the KL divergence is supported by the same subspace, as illustrated in (b). This conclusion can actually be derived from the worst-case measure given by Eq.~\ref{eq:sigmakl} (see Appendix E for proof).

On the other hand, the Wasserstein approach is capable of examining measures supported by other vector subspaces. We first ignore the constant variance term by setting $\alpha$ to zero in Eq.~\ref{eq:sigmaw}. The Wasserstein approach ``rotates'' the original support by applying linear maps to the reference measure. In the case illustrated by Fig.~\ref{fig:5}(c), essentially all measures supported by a one-dimensional vector subspace are within the scope of the approach (see Appendix F for proof). Among those measures the Wasserstein approach picks the worst one, supported by a vector subspace different from the original one. It essentially searches for the optimal transform over the entire space. In practice, we may want to account for the risk associated with the assumption of perfect correlation. This is accomplished by assigning positive value to $\alpha$, allowing the distribution to ``diffuse'' into the entire vector space as illustrated in Fig.~\ref{fig:5}(d).

It is worthwhile noting that the Wasserstein approach also has a practical advantage over the approach based on KL divergence. If we examine the worst-case variances resulting from the two approaches, Eq.~\ref{eq:sigmaw} and \ref{eq:sigmakl}, we can find that their positive definiteness is not guaranteed. This requires practitioners to carefully parametrise either approach to ensure the positive definiteness. However, under KL divergence the positive definiteness is dependent on the original covariance matrix. This makes it harder to parametrise and generalise the approach. In cases where the asset returns have time-varying correlations, one may need to switch parameters ($\theta$) to ensure a positive definite matrix. On the other hand, the Wasserstein approach only requires $B-\beta A$ to be positive-definite, independent of the covariance matrix $\Sigma$. The reference probability measure thus no longer affects the feasibility of quantifying the worst-case risk.

\subsection{Robust Portfolio Optimisation and Correlation Risk}
In modern portfolio theory, one considers $n$ securities with the excess logarithmic returns following a multivariate normally distribution, i.e. $X\sim\mathcal{N}(\mu,\Sigma)$. The standard mean-variance optimisation is formulated by
\begin{align}
&\min_a a^T\Sigma a\\
&s.t. \mu^Ta= C
\end{align}
where $a\in\mathbb{R}^n$ is the vector of portfolio weights. It can take any values assuming it is always possible to borrow or lend at the risk-free rate, and to short sell any asset. The problem is solved by introducing a Lagrange multiplier $\lambda$:
\begin{align}\label{eq:opt1}
 a^*=\frac{\lambda}{2}\Sigma^{-1}\mu
\end{align}
The optimal portfolio weight $a^*$ depends on $\lambda$. However, the Sharpe ratio of the optimal portfolio is independent of $\lambda$:
\begin{align}
\frac{a^{*T}\mu}{\sqrt{a^{*T}\Sigma a^*}}=\sqrt{\mu^T\Sigma^{-1}\mu}
\end{align}

The reference model assumes a multivariate normal distribution $\mathcal{N}(\mu,\Sigma)$. The worst-case model is an alternative measure dependent on the security positions $a$. To formulate the problem of worst-case measure, we may first express the mean-variance optimisation problem by 
\begin{align}\label{eq:mina}
\min_a E\left[(x-\mu)^Taa^T(x-\mu)-\lambda x^Ta\right]
\end{align}
where the expectation is taken under the reference measure. Taking into account the model risk, we may formulate a robust version of Eq.~\ref{eq:mina} that is consistent with literature work \cite{glasserman2014robust}:
\begin{align}\label{eq:maxm}
 \min_a \max_{Q\in\mathscr{M}} E^Q\left[(X-\mu)^Taa^T(X-\mu)-\lambda X^Ta\right]
\end{align}
where $\mathscr{M}$ is the space of alternative measures constrained by different criteria. For the approached based on the Kullback-Leibler divergence, the constraint is given by a maximum amount of relative entropy w.r.t the reference model (i.e. relative entropy budget). Under the Wasserstein approach, the constraints are given by Eq.~\ref{eq:constraints2} and \ref{eq:constraints3}.

To solve the inner problem of Eq.~\ref{eq:maxm}, we may further simplify the problem to
\begin{align}\label{eq:problemrobust}
 &\max_{Q\in\mathscr{M}} E^Q\left[(X-\mu)^Taa^T(X-\mu)-\lambda X^Ta\right]\nonumber\\
 =&
 \max_{Q\in\mathscr{M}} E^Q\left[(X-\mu-k)^Taa^T(X-\mu-k)\right]-\lambda\mu^Ta-\frac{\lambda}{4}
\end{align}
where $k$ is a vector that satisfies $a^Tk=\lambda/2$. It is noted that this is an approximation as the change of measure would also alter the mean from $\mu$ to $\mu'$. The variance should be calculated by $E^{Q(m)}\left[(X-\mu')^Taa^T (X-\mu')\right]$. However, the difference is proportional to $(\mu'-\mu)^2$ and is thus secondary on a small change of measure (i.e. $\beta\ll 1$). The solution to Eq.~\ref{eq:maxm} is also multivariate normal under both KL divergence (see Appendix G) and under the Wasserstein metric (see Appendix H). 
The two approaches result in robust MVO portfolios with different weights (up to the first order w.r.t $\theta$ or $\beta$):
\begin{align}\label{eq:robustwe}
a^*_{KL}=&\left(\frac{\lambda}{2}-\frac{\theta\lambda^3}{2}\left(1+\mu^T\Sigma^{-1}\mu\right)\right)\Sigma^{-1}\mu\nonumber\\
a^*_W=&\left(\frac{\lambda}{2}-\frac{\beta\lambda^3}{4}\mu^T\Sigma^{-1}B^{-1}\Sigma^{-1}\mu\right)\Sigma^{-1}\mu-\frac{\beta\lambda^3}{4}\left(1+\mu^T\Sigma^{-1}\mu\right)\Sigma^{-1}B^{-1}\Sigma^{-1}\mu
\end{align}
Comparing Eq.~\ref{eq:robustwe} with the standard MVO portfolio given by Eq.~\ref{eq:opt1}, we can see that the robust MVO portfolios provide first-order corrections, resulting in more conservative asset allocation in general.

\begin{comment}
This reduces the outer part of Eq.~\ref{eq:maxm} to
\begin{align}
&\min_a a^T(I-2\theta\Sigma A)^{-1}\Sigma a-\lambda a^T(\mu-2\theta(I-2\theta\Sigma A)^{-1}\Sigma a)\nonumber\\
=&\min_a a^T\Sigma a-\lambda a^T\mu+2\theta a^T\Sigma a[\lambda+(1+2\theta \lambda) a^T\Sigma a)]+O(\theta^2)
\end{align}
under the KL divergence, and 
\begin{align}
&\min_a a^T(I-\beta B^{-1}A)^{-1}\Sigma(I-\beta AB^{-1})^{-1}a-\lambda a^T\left[\mu-\beta(I-\beta B^{-1} A)^{-1}B^{-1} a\right]\nonumber\\
=&\min_a a^T\Sigma a-\lambda a^T\mu+\beta \left(2a^TB^{-1}aa^T\Sigma a%+a^T\Sigma a a^T\right]
+\lambda a^TB^{-1}a\right)+O(\beta^2)
\end{align}
under the Wasserstein approach. 
Ignoring the higher order terms, one may solve the minimisation problems via linearisation:
\begin{align}\label{akl}
a^*=\left(\frac{\lambda}{2}-\theta\left[2\lambda+(2+4\theta\lambda)\mu^T\Sigma^{-1}\mu\right]\right)\Sigma^{-1}\mu=c\Sigma^{-1}\mu
\end{align}
under the KL divergence, and 
\begin{align}\label{aw}
a^*=&\left(\frac{\lambda}{2}-2\beta \mu^T\Sigma^{-1}B^{-1}\Sigma^{-1}\mu\right)\Sigma^{-1}\mu
-\beta\left(2\mu^T\Sigma^{-1}\mu+\lambda\right)\Sigma^{-1}B^{-1}\Sigma^{-1}\mu
\end{align}
\end{comment}

Despite of being more conservative, $a^*_{KL}$ is in fact parallel to the standard MVO portfolio $a^*$. As a result, the robust MVO portfolio does not change the relative weights of component assets. In fact, all the weights are reduced by the same proportion ($c<1$) to account for model risk. This is, however, inappropriately accounts for the correlation risk. For example, two highly-correlated assets have extremely high weights in the nominal MVO portfolio. Because of the correlation risk, we would expect the robust MVO portfolio to assign them lower weights relative to other assets. 
This is the case for $a^*_W$. In fact, $a^*_W$ not only reduces the overall portfolio weights in order to be more conservative, but also adjusts the relative weights of component assets for a less extreme allocation. One may notice that the term inside the bracket of the expression for $a^*_W$ is a square matrix (see Eq.~\ref{eq:robustwe}), which serves to linearly transform the vector of portfolio weights. By adjusting their relative weights, Eq.~\ref{eq:wweights} correctly accounts for the correlation risk (see Appendix H for details). 

The robust optimal portfolio parametrised by $\lambda$ allows us to plot the robust capital market line (CML). Unlike the standard CML, it is no longer a straight line and the Sharpe ratio is now dependent on $\lambda$. 

\begin{figure}[H]
\begin{center}
\includegraphics[width=.45\textwidth]{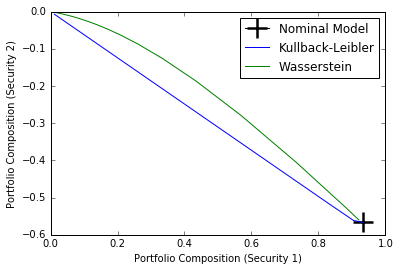}
\includegraphics[width=.45\textwidth]{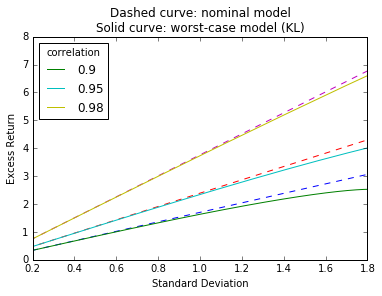}
\includegraphics[width=.45\textwidth]{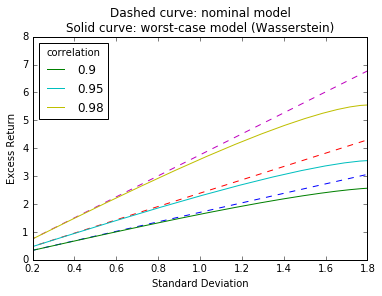}
\renewcommand{\figurename}{Fig}
\caption{The normalised optimal composition of a portfolio consisting of two securities, calculated by $a^*$ divided by $\lambda/2$. The normalised optimal composition under the reference model is give by a constant vector $\Sigma^{-1}\mu$, while those under the worst-case models are dependent on $\lambda$. In particular, the Kullback-Leibler approach reduces both compositions proportionally, while the Wasserstein approach reduces compositions in a nonlinear way.}\label{00}
\end{center}
\end{figure}

\begin{figure}[H]
\begin{center}
\includegraphics[width=.45\textwidth]{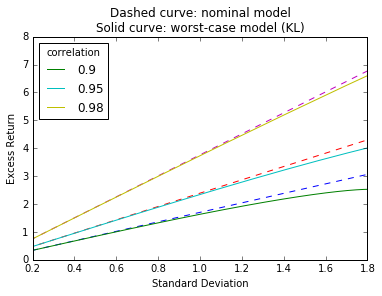}
\includegraphics[width=.45\textwidth]{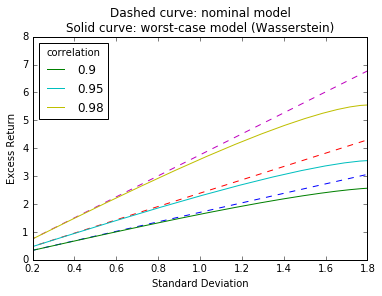}
\renewcommand{\figurename}{Fig}
\caption{Robust capital market lines (CMLs) using (a) the Kullback-Leibler divergence and (b) the Wasserstein approach.}\label{01}
\end{center}
\end{figure}

Under the reference model, the optimal composition of a portfolio is given by $\lambda\Sigma^{-1}\mu/2$. The proportionality of this solution suggests that we should double the weights if the expected excess return doubles. However, this may end up with excessive risk due to increase of leverage. Model risk is the major source of risks here, as we are unsure if the expected excess return and the covariance matrix correctly reflects the return distribution in the future (for a given holding period). Since higher leverage implies more severe model risk, increasing leverage proportionally is in fact sub-optimal under the worst-case model. 

Eq.~\ref{eq:robustwe}, on the other hand, provides the optimal solutions under the respective model risk approaches. The robustness of these solutions allow the practitioners to allocate assets in a safer way. It is shown in Fig.~\ref{00} that the normalised optimal compositions reduce with $\lambda$. This is because a larger $\lambda$ indicates higher leverage, and hence the optimal composition is reduced further away from that of the reference model. %The difference between the two model risk approaches lie in the 
The normalised optimal compositions approach zero on the increase of $\lambda$. 
In Fig.~\ref{00}, the compositions of both securities get reduced proportionally under the KL approach. Using the Wasserstein approach, on the other hand, allows the compositions to move in a non-parallel way. 

In this example, we have two highly correlated ($\rho=0.5$) stocks but with very different expected excess returns (Stock 1 $0.65$ and Stock 2 $-0.1$). Because of the high correlation we can profit from taking the spread (long Stock 1 and short Stock 2). Under the reference model, taking the spread of a highly correlated pair does not add too much risk. However, the true risk could be underestimated due to the existence of model risk. The spread is more sensitive to model risk than an overall long position, and thus requires reduction when optimising with model risk. This point is well reflected by the non-linearity of the capital market line under the Wasserstein approach, showing sub-linear increase in excess return as risk (standard deviation) increases. We reduce the position of the spread more than the long position of Stock 1 (or the overall long position). In the KL approach, however, we reduce the spread position and the overall long position at the same pace.

The effect of robust optimality under the worst-case model is most significant when the reference model is close to having a low-dimensional support. A low-dimensional support means that the covariance matrix does not have the full rank. Put in a practical way, there exists a risk-free portfolio with non-zero compositions in risky assets. In this case, there is arbitrage opportunity that has close-to-zero risk but high excess returns. The optimal portfolio under the reference model could be unrealistically optimistic, i.e. the arbitrage opportunity might disappear in the face of model risk.

Fig.~\ref{00} illustrates an example of two securities with a high correlation. Under the reference model, the Sharpe ratio (slope of the excess return vs risk line) increases quickly with the correlation coefficient, demonstrated by the dashed lines in Fig.~\ref{01}. This results from taking excessive positions in the spread (long the one with higher Sharpe ratio and short the other). It is clear from Fig.~\ref{01} that the approach based on the Kullback-Leibler divergence cannot solve this issue systematically. In fact, when the correlation increases, the capital market line under the worst-case model is even closer to the nominal one. On the other hand, the Wasserstein approach does provide a more plausible adjustment. The robust capital market line given by the Wasserstein approach deviates more from the nominal straight line on an increasing correlation.

This difference is a direct result in their capabilities of altering the support of the reference measure. The KL approach cannot alter the support. So a spurious arbitrage relation under the reference measure may persist under the worst-case measure. On the other hand, the Wasserstein %search for a support for the worst-case measure. It 
approach breaks the ostensible arbitrage opportunity by transforming the support to a different vector subspace.

\subsection{Model Risk in Dynamic Hedging}

The hedging error is measured by the absolute profit-and-loss (PnL) of a dynamically hedged option until its maturity. Using the Black-Scholes model as the reference model, the hedging risk decreases with the hedging frequency. Ideally if hedging is done continuously, then the hedging error is zero almost surely. This is true even under alternative measures, as long as they are equivalent to the reference model. %, as the support does not change. 
The underlying reason is that the quadratic variation does not change under all equivalent measures. In fact, if we consider a geometric Brownian motion:
\begin{align}
dS_t=\mu S_tdt+\sigma S_tdW_t
\end{align}
The quadratic variation $[\ln S]_t=\int_0^t \sigma^2_sds$ almost surely. Therefore the equation holds under all equivalent measures. Given the Black-Scholes price of an option $C_t=C(t,S_t)$, the PnL of a continuously hedged portfolio between time $0$ and $T$ is
\begin{align}\label{eq:pnl}
&\int_0^TdC_t-\int_0^T\frac{\partial C_t}{\partial S_t}dS_t\nonumber\\
=&\int_0^T\left(\frac{\partial C_t}{\partial t}dt+\frac{S_t^2}{2}\frac{\partial^2 C_t}{\partial  S_t^2}d[\ln S]_t\right)
%=&\int_0^T\left(\frac{\partial C_t}{\partial t}+\frac{1}{2}\frac{\partial^2 C_t}{\partial S_t^2}\sigma^2_tS_t^2\right)dt\nonumber\\
=0
\end{align}
where the last equality results from the Black-Scholes partial differential equation. 

\begin{figure}[H]
\begin{center}
\includegraphics[width=.45\textwidth]{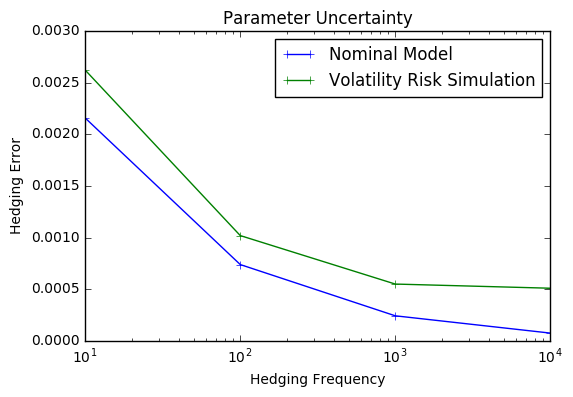}
\includegraphics[width=.45\textwidth]{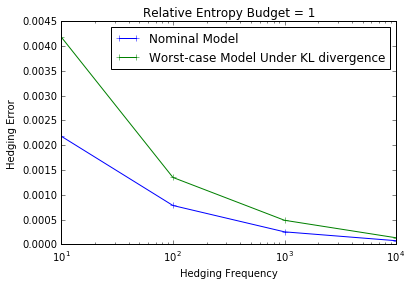}
\renewcommand{\figurename}{Fig}
\caption{(a) Worst-case hedging risk under the KL divergence, and (b) hedging risk simulated by randomly sampling volatilities.}\label{hedge}
\end{center}
\end{figure}

Since any $f$-divergence is only capable of searching over equivalent alternative measures, the worst-case hedging error given by these approaches has to be zero on continuous hedging frequency. 
One can image that as hedging frequency increases, the worst-case hedging risk decreases towards zero (Fig.~\ref{hedge}(b)). This is, however, inconsistent with practitioners' demand for risk management. In fact, if the volatility of the underlying asset differs from the nominal volatility, then Eq.~\ref{eq:pnl} no longer holds. Such volatility uncertainty is a major source of hedging risk, and thus has to be measured and managed properly. The most straightforward way of doing that is to assume a distribution of volatility, and then run a Monte Carlo simulation to quantify the hedging error (Fig.~\ref{hedge}(a)).

Despite of its simplicity, volatility sampling is a parametric approach, for it is only capable of generating alternative Black-Scholes models with different parameter. This approach cannot account for alternatives such as local volatility models or stochastic volatility models. This calls for a non-parametric approach relying on the formulation given in Eq.~\ref{eq:basic}. 

%Sampling volatility limits the type of alternative measures in a small parametric subset, while in reality the measure can be of a different kind. 

%For example, the real measure (on the return process) could be a mean-reverting OU process instead of a Brownian motion, which is inclusive by the KL divergence but exclusive by volatility sampling. Moreover, if the real measure requires complex dynamics of the volatility, following a local volatility or a stochastic volatility model for instance, then neither approaches can involve such an alternative measure.

\begin{figure}[H]
\begin{center}
\includegraphics[width=.6\textwidth]{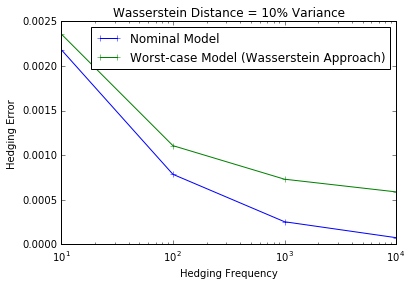}
\renewcommand{\figurename}{Fig}
\caption{(a) Worst-case hedging risk under the Wasserstein approach.}\label{hedgew}
\end{center}
\end{figure}

We have already seen that using approaches based on $f$-divergence one cannot correctly quantify the hedging risk. 
The Wasserstein approach, on the other hand, does not have this issue, for it is capable of searching over non-equivalent measures. 
Using Monte Carlo simulation, we obtain the worst-case hedging risk under the Wasserstein approach (see Fig. \ref{hedgew}). Compared to the approach based on Kullback-Leibler divergence (Fig.~\ref{hedge}(b)), the hedging risk given by the Wasserstein approach is more consistent with the simulated results using volatility sampling (Fig.~\ref{hedge}(a)). In the limit of continuous hedging, the Wasserstein approach results in a worst-case risk slightly higher than volatility sampling, for it may involve jumps that cannot be hedged.

\begin{figure}[H]
\begin{center}
\includegraphics[width=.4\textwidth]{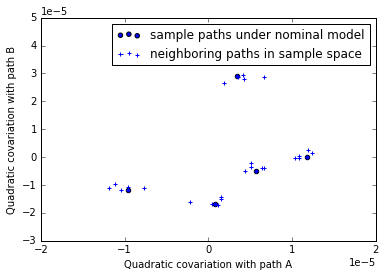}
\includegraphics[width=.45\textwidth]{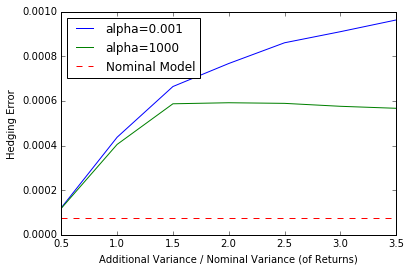}
\renewcommand{\figurename}{Fig}
\caption{(a) Sample paths generated for the Wasserstein approach, (b) convergence of the worst-case hedging risk.}\label{hedgewpts}
\end{center}
\end{figure}

In practice, the Wasserstein approach requires some tricks as fully sampling the infinite-dimensional path space is impossible. Therefore only paths close to the sampled paths (under the reference measure) are sampled, as the importance of an alternative path decays exponentially with its distance to these sampled paths. This point is shown in Fig. \ref{hedgewpts}(a), in which the alternative paths are illustrated by the crosses close to the nominal sampled paths (dots). By increasing the average distance of the alternative paths to the nominal paths, the hedging risk is increased until convergence (Fig. \ref{hedgewpts}(b)).

Here we list the procedure of the Monte Carlo simulation described in the last paragraph:\\
1. create $N$ sample paths from the reference model\\
2. For each sample paths, create $M$ sample paths by deviating $X_t$ by a normally distributed random variable $\mathcal{N}(0,\sigma^2)$\\
3. collect all $MN$ sample paths and the original $N$ paths, we have $N(M+1)$ points in the path space. Calculate the hedging error for each of the $N(M+1)$ paths.\\
4. Apply Eq.~\ref{eq:main} to calculate the worst-case probability of each path where $d(X,Y)=[X-Y]$.\\
5. To find the (worst-case) hedging risk, we average the hedging errors of all $N(M+1)$ paths, weighted by their worst-case probabilities. \\
6. Repeat steps 2-5 with a larger $\sigma^2$. Continue to increase the deviation until the calculated hedging risk converges (Fig.~\ref{hedgewpts}(b)).

\section{Conclusion}
Non-parametric approaches to model risk measurement are theoretically sound and practically feasible. Adopting the Wasserstein distance allows us to further extend the range of legitimate measures from merely the absolutely continuous ones. This Wasserstein approach roots in optimal transport theory and is well suited for the adversary interpretation of model risk. In particular, it specifies the economic reality of the fictitious adversary with the capacity of parametrising the actual market structure. The Wasserstein approach may result in the worst-case model that is more robust, in the sense that it is no longer restricted by the support of the reference measure. This is especially useful when the reference measure is supported only by a subspace (for instance the volatility of a diffusion process or the prices of perfectly correlated assets). This approach has additional practical advantage due to its ability of guaranteeing integrability. 

To further illustrate the Wasserstein approach, we presented four applications ranging from single-asset variance risk and hedging risk to the multi-asset allocation problem. All the applications are connected in the sense that their reference measures are (or close to) supported by merely a subspace. In the example of single-asset variance risk, we look at the limit of small variance, i.e. when the time to maturity is close to zero (or the volatility close to zero). The Wasserstein approach is capable of jumping out of the family of diffusion processes, and accounts for the possibility of jumps. In the application of portfolio variance risk, the Wasserstein approach provides us with worst-case measure induced by a linear map, thus altering the support. Its advantage of dealing with multi-asset problems is even more apparent when treating the asset allocation problem, in which the Wasserstein approach accounts for the correlation risk. This approach results in a robust mean-variance optimal portfolio that adjusts the relative weights of the assets according to their correlations. It produces a curved capital allocation line, with the Sharpe ratio reduced by a larger amount on a higher standard deviation or a higher asset correlation. The final application is related to the hedging risk of a vanilla option. $f$-divergence is incapable of quantifying the risk associated with a continuously hedged position because its profit-and-loss is zero almost surely. The Wasserstein approach, on the other hand, leads to a positive hedging error and therefore a more realistic assessment of model risk. %in consistency with the simulation result. 
In conclusion, the Wasserstein approach provides a useful tool to practitioners who aim to manage risks and optimize positions accounting for model ambiguity.

\section{Appendix}

\subsection{A. Derivation of Eq.~\ref{eq:ystar}}
  
In this part, we derive the solution Eq.~\ref{eq:ystar} to the problem expressed by Eq.~\ref{eq:prob}-\ref{eq:constraints2}. 
For simplicity, we denote the transition density $p_{Y|X}(y|x)$ by $\gamma_x(y):=\gamma(x,y)/p(x)$. This transforms the problem into
\begin{align}\label{eq:proof1}
\sup_{\gamma_x\in\Gamma}&\int_{\Omega}p(x)\left[\int_{\Omega} \gamma_x(y)V(y)dy\right]dx\\
s.t.
&\int_{\Omega}p(x)\left[\int_{\Omega}  \gamma_x(y)c(x,y)dy\right]dx\leq\eta\nonumber
%&\int_{\Omega} \gamma_x(y)dy=1,~\forall x\in\Omega\nonumber
\end{align}
where $\Gamma$ is the space of probability density functions. The Karush-Kuhn-Tucker (KKT) condition in convex optimisation ensures the existence of a KKT multiplier $\lambda$ such that the solution to Eq.~\ref{eq:proof1} also solves
\begin{align}\label{eq:proof12}
\sup_{\gamma_x\in\Gamma}&\int_{\Omega}p(x)\left\{\int_{\Omega} \gamma_x(y)\left[V(y)-\lambda c(x,y)\right]dy\right\}dx
\end{align}
The solution to Eq.~\ref{eq:proof12} is a $\delta$-function transition density $\gamma_x^*(y)=\delta\left(y-y^*(x)\right)$, resulting in a transportation plan
\begin{align}\label{eq:plan}
\gamma^*(x,y)=p(x)\delta\left(y-y^*(x)\right)
\end{align}
where
\begin{align}\label{eq:map}
y^*(x)=\arg\max_{y\in\Omega}\left[V(y)-\lambda c(x,y)\right]
\end{align}
The solution to the model risk problem is expressed either by a transportation plan (Eq.~\ref{eq:plan}) or a transportation map (Eq.~\ref{eq:map}). It is noted that $\lambda=0$ is a trivial case that we will not consider. To be consistent with the main result Eq.~\ref{eq:main}, we replace $\lambda$ by its inverse $\beta=\lambda^{-1}$:
\begin{align}
y^*(x)=\arg\max_{y\in\Omega}\left[V(y)-\frac{c(x,y)}{\beta}\right]
\end{align}

\subsection{B. Derivation of Eq.~\ref{eq:main} and \ref{eq:8} }

Eq.~\ref{eq:main} is the solution of the problem formulated by Eq.~\ref{eq:prob}-\ref{eq:constraints2} plus the additional entropy constraint Eq.~\ref{eq:constraints3}. As in Appendix A, 
%To solve the problem formulated in Eq.~\ref{eq:prob}, 
we introduce KKT multipliers $\lambda$ and $\alpha$. This converts the original constrained supremum problem to the following dual problem (same as in Appendix A we denote the transition density by $\gamma_x(y)$):
\begin{align} \label{eq:app1}
&\inf_{\beta, \theta\in\mathbb{R}^+}\sup_{\gamma}%\mathbb{E}_X\left(
\int_{\Omega\times\Omega}\gamma(x,y)\left(V(y)- \lambda\left[c(x,y)-\eta\right]-\alpha\left[\ln \gamma(x,y)-\mu\right]\right)dxdy%+\frac{\eta}{\beta}+\alpha\left(\mu-\mathbb{E}\left[\ln p(X)\right]\right)
\\=&
\inf_{\beta, \theta\in\mathbb{R}^+}\left(%\sup_{p_{Y|X}(y|x)}%\mathbb{E}_X\left(
\int_\Omega p(x)dx
\left[\sup_{\gamma_x}\int_\Omega \gamma_x(y)\left(V(y)- \lambda c(x,y)-\alpha\ln\gamma_x(y)\right)dy\right]\right.\nonumber\\
&~~~~~~~~~~~~\left.+\lambda\eta+\alpha\left[\mu-\int_\Omega\ln p(x)dx\right]\right)\nonumber
\end{align}
Same as the relative entropy approach proposed by Glasserman and Xu \cite{glasserman2014robust}, we derive a closed-form solution to the inner part of the problem:
\begin{align}\label{eq:inner}
\sup_{\gamma_x}\int_\Omega \gamma_x(y)\left(V(y)- \lambda c(x,y)-\alpha\ln\gamma_x(y)\right)dy
%\sup_{p_{Y|X}(y|x)}\mathbb{E}\left(V(Y)- \frac{d(X,Y)}{\beta }-\alpha\ln f(Y|X)\right) 
\end{align}
It is noted that Eq.~\ref{eq:inner} asks for the supremum w.r.t the density function $p_{x}$ for a given $x\in\Omega$. %Fortunately, the rest of Eq.~\ref{eq:app1} is independent of $p_{Y|X}$. 
The solution to this problem is given by (for consistency we replace $\lambda$ by its inverse $\gamma$):
%Solving the supermium problem for the first term gives
\begin{align}\label{eq:gammaxy}
\gamma_x^*(y)=\frac{\exp\left(\frac{V(y)}{\alpha}-\frac{c(x,y)}{\alpha\beta }\right)}{\int_\Omega\exp\left(\frac{V(z)}{\alpha}-\frac{c(x,z)}{\alpha\beta }\right)dz}%\right|X=x\right]}
\end{align}
%which is the solution for the problem of restricted risk source.
The worst-case probability density function is the marginal distribution of $y$, induced by the transition density function $\gamma^*_x(y)$:
\begin{align}
p^*(y)=&\int_\Omega p(x)\gamma_x^*(y)dx\nonumber\\
=&\int_\Omega dx\frac{p(x)\exp\left(\frac{V(y)}{\alpha}-\frac{c(x,y)}{\alpha\beta }\right)}{\int_\Omega\exp\left(\frac{V(z)}{\alpha}-\frac{c(x,z)}{\alpha\beta }\right)dz}%\right|X=x\right]}
%=&\mathbb{E}_X\left(\frac{\exp\left(\frac{V(y)}{\alpha}-\frac{c(x,y)}{\alpha\beta }\right)}{\mathbb{E}_Y\left[\left.\exp\left(\frac{V(y)}{\alpha}-\frac{c(x,y)}{\alpha\beta }\right)\right|X=x\right]}\right)
\end{align}

Eq.~\ref{eq:8} is derived in a similar way. Since we lift the entropy constraint Eq.~\ref{eq:constraints3} into a relative entropy constraint Eq.~\ref{eq:constraints4}, the inner problem Eq.~\ref{eq:inner} requires slight modification:
\begin{align}\label{eq:inner1}
\sup_{\gamma_x}\int_\Omega \gamma_x(y)\left(V(y)- \lambda c(x,y)-\alpha\ln\frac{\gamma_x(y)}{q_0(y)}\right)dy
\end{align}
This problem has the same formulation as the supremum problem given in Glasserman and Xu's work, and therefore shares the same solution
\begin{align}\label{eq:gammaxy1}
\gamma_x^*(y)=\frac{q_0(y)\exp\left(\frac{V(y)}{\alpha}-\frac{c(x,y)}{\alpha\beta }\right)}{\int_\Omega q_0(z)\exp\left(\frac{V(z)}{\alpha}-\frac{c(x,z)}{\alpha\beta }\right)dz}%\right|X=x\right]}
\end{align}
This equation differs from Eq.~\ref{eq:gammaxy} merely by a prior distribution $q_0$. It takes Eq.~\ref{eq:gammaxy} as its special case where $q_0$ is a uniform distribution. Marginalizing the transition density Eq.~\ref{eq:gammaxy1} gives the worst-case distribution shown in Eq.~\ref{eq:8}.
%\subsection{C. Derivation of Eq.~\ref{}}

\subsection{C. Jump Risk and Variance Risk}
Under a diffusive model, the logarithmic return of an asset follows a normal distribution with mean of $\mu T$ and variance of $\sigma^2T$, where $\sigma$ is the volatility and $T$ is the time to maturity, and the drift coefficient of this process is assumed to be $\mu+\sigma^2/2$. The probability density function of the return $x$ is
\begin{align}
p(x)=
\frac{1}{\sqrt{2\pi}\sigma}e^{-\left(\frac{(x-\mu T)^2}{2\sigma^2T}\right)^2}
\end{align}
Applying Eq.~\ref{eq:main}, one may obtain the probability density function of the worst-case measure, assuming a linear loss function $V(x)=x$ and a quadratic transportation cost function $c(x,y)=(x-y)^2$,
\begin{align}\label{eq:varianceqy}
q^*(y)\propto
&\int_\Omega \left[p(x)\left.
\exp\left(\frac{y}{\alpha}-\frac{(x-y)^2}{\alpha\beta}\right)\right/
\exp\left(\frac{x}{\alpha}\right)\right]dx\nonumber\\
=&\int_\Omega \exp\left(\frac{y-x}{\alpha}-\frac{(x-y)^2}{\alpha\beta}-\frac{(x-\mu T)^2}{2\sigma^2T}\right)dx\nonumber\\
\propto&\exp\left(-\frac{\left(y-\mu T-\beta/2\right)^2}{2\sigma^2T+\alpha\beta}\right)
\end{align}
Unlike the result given by the KL divergence, Eq.~\ref{eq:varianceqy} not only shifts the mean the distribution but also enlarges the variance as a result of additional uncertainty. On $\sigma\to0$, the worst-case measure is no longer a Dirac measure, showing consideration of jump risks:
\begin{align}
\lim_{\sigma\to0}q^*(y)\propto\exp\left(-\frac{\left(y-\mu T-\beta/2\right)^2}{\alpha\beta}\right)
\end{align} 
This gives Eq.~\ref{eq:linearex}. Alternatively, one may first derive Eq.~\ref{eq:jump} followed by substituting $V(x)=x$ to get Eq.~\ref{eq:linearex}. Eq.~\ref{eq:jump} is derived by substituting $p(x)=\delta(x-\mu T)$ into Eq.~\ref{eq:main}:
\begin{align}
q^*(y)=&\int_\Omega  \delta(x-\mu T)\frac{\exp\left(\frac{V(y)}{\alpha}-\frac{(x-y)^2}{\alpha\beta }\right)}{\int_\Omega\exp\left(\frac{V(z)}{\alpha}-\frac{(x-z)^2}{\alpha\beta }\right)dz}dx\\
=&\frac{\exp\left(\frac{V(y)}{\alpha}-\frac{(y-\mu T)^2}{\alpha\beta }\right)}{\int_\Omega\exp\left(\frac{V(z)}{\alpha}-\frac{(z-\mu T)^2}{\alpha\beta }\right)dz}
\end{align}

Now we adopt a quadratic type of loss function, $V(x)=(x-\mu T)^2$, following a procedure similar to Eq.~\ref{eq:varianceqy} we get
\begin{align}\label{eq:measureq}
q^*(y)\propto\exp\left(-\frac{\left(y-\mu T\right)^2}{\frac{2\sigma^2T}{(1-\beta)^2}+\frac{\alpha\beta}{(1-\beta)}}\right)
\end{align}
the variance of the worst-case measure is
\begin{align}\label{eq:maxvar}
\sigma_W^2T=\frac{\sigma^2T}{(1-\beta)^2}+\frac{\alpha\beta}{2(1-\beta)}
\end{align}
as provided in Eq.~\ref{eq:swt}.
We may verify that the measure $Q^*$ given by Eq.~\ref{eq:measureq} does provide the largest variance among all the legitimate alternative measures. In fact, the variance of $x$ under $Q^*$ is
\begin{align}
E^{Q^*}\left[\left(x-E^{Q^*}(x)\right)^2\right]=E^{Q^*}\left[\left(x-\mu T\right)^2\right]
\end{align}
According to the definition of the worst-case model, for all $Q\in\mathscr{M}$ (the space of legitimate alternative measures) we have
\begin{align}
E^{Q^*}\left[\left(x-\mu T\right)^2\right]\geq& E^{Q}\left[\left(x-\mu T\right)^2\right]\\
=&E^{Q}\left[\left(x-E^{Q}(x)\right)^2\right]+\left(E^{Q}(x)-\mu T\right)^2\\
\geq&E^{Q}\left[\left(x-E^{Q}(x)\right)^2\right]
\end{align}
This confirms that Eq.~\ref{eq:maxvar} is indeed the worst-case (maximum) variance.

\subsection{D. Worst-case Portfolio Variance}
To find the portfolio variance under the worst-case scenario, we need to formulate the problem using Eq.~\ref{eq:basic} with a loss (target) function given by Eq.~\ref{eq:xAx}. The worst-case measure may be evaluated by substituting the loss function into Eq.~\ref{eq:main}. In this section we will show the calculation step by step. First, we need to specify the transport cost function $c(x,y)$ as the inner product introduced in Eq.~\ref{eq:innerpro}:
\begin{align}
c(x,y)=||y-x||^2=(y-x)^TB(y-x)
\end{align}
Then we evaluate the following part in Eq.~\ref{eq:main}:
\begin{align}\label{eq:expvd}
&\exp\left(\frac{V(y)}{\alpha}-\frac{c(x,y)}{\alpha\beta}\right)\nonumber\\
=&\exp\left(\frac{y^TAy}{\alpha}-\frac{(y-x)^TB(y-x)}{\alpha\beta}\right)\nonumber\\
=&
\exp\left(\frac{1}{\alpha\beta}x^TB\left((B-\beta A)^{-1}-I\right)Bx\nonumber\right.\\
&~~~~~~~~\left.-\frac{1}{\alpha\beta}\left(y-(B-\beta A)^{-1}Bx\right)^T\left(B-\beta A\right)\left(y-(B-\beta A)^{-1}Bx\right)\right)
\end{align}
Remember that both $A$ and $B$ are symmetric, positive-definite matrices. Fixing $x$, Eq.~\ref{eq:expvd} is proportional to the probability density function of a multivariate normal variable $Y$, with its mean and covariance matrix 
\begin{samepage}
\begin{align}
\mathsf{E}(Y)=&(B-\beta A)^{-1}Bx\\
\Sigma(Y)=&\frac{\alpha\beta}{2}(B-\beta A)^{-1}
\end{align}
\end{samepage}
This means that after normalization w.r.t $y$, Eq.~\ref{eq:expvd} gives exactly the probability density function of $Y$. We may write this down explicitly by noting that $y$ lives in the $n$-dimensional vector space, i.e. $\Omega=\mathcal{V}$:
\begin{align}\label{eq:normalise}
&\frac{\exp\left(\frac{V(y)}{\alpha}-\frac{c(x,y)}{\alpha\beta }\right)}{\int_{\mathcal{V}}\exp\left(\frac{V(y)}{\alpha}-\frac{c(x,y)}{\alpha\beta }\right)dy}\\
=&
(2\pi)^{-\frac{n}{2}}\sqrt{\frac{\alpha\beta}{2}|B-\beta A|}
\exp\left(-\left(y-(B-\beta A)^{-1}Bx\right)^T\frac{\left(B-\beta A\right)}{\alpha\beta}\left(y-(B-\beta A)^{-1}Bx\right)\right)\nonumber
\end{align}
Now we need to evaluate the product of Eq.~\ref{eq:normalise} and the nominal distribution $p(x)$. The nominal distribution is multivariate normal with mean $\mu$ and covariance matrix $\Sigma$:
\begin{align}
p(x)=\frac{(2\pi)^{-\frac{n}{2}}}{|\Sigma|}\exp\left(-\frac{1}{2}(x-\mu)^T\Sigma^{-1}(x-\mu)\right)
\end{align}
The product contains many terms of $x$ and $y$. One may re-arrange the terms to isolate quadratic and linear terms of $x$:
\begin{align}\label{eq:pxyvar}
&\frac{p(x)\exp\left(\frac{V(y)}{\alpha}-\frac{c(x,y)}{\alpha\beta }\right)}{\int_{\mathcal{V}}\exp\left(\frac{V(y)}{\alpha}-\frac{c(x,y)}{\alpha\beta }\right)dy}\nonumber\\
%=&(2\pi)^{-\frac{n}{2}}\sqrt{\frac{\alpha\beta}{2}\frac{|B-\beta A|}{|\Sigma|}}
\propto&
\exp\left(-\frac{1}{\alpha\beta}\left(y-(B-\beta A)^{-1}Bx\right)^T\left(B-\beta A\right)\left(y-(B-\beta A)^{-1}Bx\right)\right.\nonumber\\
&~~~~~~~~~~~~~~~~~~~~~~~~~~~~~~~~~~~~~~~~~~~~~~~~~~~~~~~~~~~~~~\left.-\frac{1}{2}(x-\mu)^T\Sigma^{-1}(x-\mu)\right)\nonumber\\
%&~~~~~~~~~~~~~~~~~~~~~~~~~~~~~~~~~~~~\left.-\frac{1}{2}(x-\mu)^T\Sigma^{-1}(x-\mu)\right)\nonumber\\
\propto&\exp\left(-\frac{1}{\alpha\beta}\left[(x-K\mu-Ly)^TM(x-K\mu-Ly)-
(K\mu+Ly)^TM(K\mu+Ly)\right]\right.\nonumber\\
&~~~~~~~~~~~~~~~~~~~~~~~~~~~~~~~~~~~~~~~~~~~~~~~~~~~~~~~~~~~~~~~~~~~~~
\left.-\frac{1}{\alpha\beta}y^T(B-\beta A)y\right)
\end{align}
where 
\begin{align}
\begin{aligned}
M:=&B(B-\beta A)^{-1}B+\frac{\alpha\beta}{2}\Sigma^{-1}\\
K:=&\frac{\alpha\beta}{2}M^{-1}\Sigma^{-1}\\
L:=&M^{-1}B
\end{aligned}
\end{align}
Fixing $y$, Eq.~\ref{eq:pxyvar} is proportional to the probability density function of a multivariate normal variable $X$ where
\begin{align}
\mathsf{E}(X)=&K\mu+Ly\\
\Sigma(X)=&\frac{\alpha\beta}{2}M^{-1}
\end{align}
The following integral
\begin{align}
\int_{\mathcal{V}}\exp\left(-\frac{1}{\alpha\beta}(x-K\mu-Ly)^TM(x-K\mu-Ly)\right)dy
=\frac{2}{\alpha\beta}(2\pi)^{-\frac{n}{2}}|M|^{-1}
\end{align}
is constant irrespective of $y$. Integrating Eq.~\ref{eq:pxyvar} over x gives the worst-case probability density function $q^*(y)$: 

\begin{align}\label{eq:fullcal}
q^*(y)=&\int_\mathcal{V} dx\frac{p(x)\exp\left(\frac{V(y)}{\alpha}-\frac{c(x,y)}{\alpha\beta }\right)}{\int_\Omega\exp\left(\frac{V(y)}{\alpha}-\frac{c(x,y)}{\alpha\beta }\right)dy}\nonumber\\
\propto&\int_\mathcal{V}\exp\left(-\frac{1}{\alpha\beta}(x-K\mu-Ly)^TM(x-K\mu-Ly)\right)dx\nonumber\\
&\times \exp\left[\frac{1}{\alpha\beta}
(K\mu+Ly)^TM(K\mu+Ly)-\frac{1}{\alpha\beta}y^T(B-\beta A)y\right]\nonumber\\
\propto&\exp\left[\frac{1}{\alpha\beta}
(K\mu+Ly)^TM(K\mu+Ly)-\frac{1}{\alpha\beta}y^T(B-\beta A)y\right]\nonumber\\
%\exp\left[\frac{1}{\alpha\beta}\left(\frac{\alpha\beta}{2}M^{-1}\Sigma^{-1}\mu+M^{-1}By\right)^TM\left(\frac{\alpha\beta}{2}M^{-1}\Sigma^{-1}\mu+M^{-1}By\right)\right.\nonumber\\&~~~~~~~~~~~~~~~~~~~~~~~~~~~~~~~~~~~~~~~~~~~~~~~~~~~~~~~~~~~~~~~~~~~~~\left.-\frac{1}{\alpha\beta}y^T(B-\beta A)y\right]\nonumber\\
=&\exp\left[\frac{1}{\alpha\beta}\left( \left(\frac{\alpha\beta}{2}\Sigma^{-1}\mu+By\right)^TM^{-1}\left(\frac{\alpha\beta}{2}\Sigma^{-1}\mu+By\right)- y^T(B-\beta A)y\right)\right]\nonumber\\
\propto&\exp\left[-\frac{1}{2}\left(y-B(B-\beta A)^{-1}\mu\right)^T 
%\left((B-\beta A)^{-1}B\Sigma B(B-\beta A)^{-1}+\frac{\alpha\beta}{2}(B-\beta A)^{-1}\right)^{-1}\right.\nonumber\\&~~~~~~~~~~~~~~~~~~~~~~~~~~~~~~~~~~~~~~~~~~~~~~~~~~~~~~~~~~~~~~~~~~~~~~~~~~~~~~~~~~~~~~~~\cdot
\Sigma^{*-1}\left(y-B(B-\beta A)^{-1}\mu\right)\right]
\end{align}
where 
\begin{align*}
\Sigma^{*-1}=&\frac{2}{\alpha\beta}\left(B^TM^{-1}B-(B-\beta A)\right)\\
=&\frac{2}{\alpha\beta}\left(\left((B-\beta A)^{-1}+\frac{\alpha\beta}{2}B^{-1}\Sigma^{-1}B^{-1}\right)^{-1}-(B-\beta A)\right)\\
=&\frac{2}{\alpha\beta}\left(\left((B-\beta A)^{-1}+\frac{\alpha\beta}{2}B^{-1}\Sigma^{-1}B^{-1}\right)^{-1}\left(I-\left(I+\frac{\alpha\beta}{2}B^{-1}\Sigma^{-1}B^{-1}(B-\beta A)\right)\right)\right)\\
=&\left((B-\beta A)^{-1}+\frac{\alpha\beta}{2}B^{-1}\Sigma^{-1}B^{-1}\right)^{-1}\left((B-\beta A)^{-1}B\Sigma B\right)^{-1}\\
=&\left((B-\beta A)^{-1}B\Sigma B(B-\beta A)^{-1}+\frac{\alpha\beta}{2}(B-\beta A)^{-1}\right)^{-1}\\
\end{align*}

Eq.~\ref{eq:fullcal} shows that the worst-case distribution is still multivariate normal. The vector of means and the covariance matrix are given respectively by
\begin{align}\label{eq:wvarresult}
\mu_W=&(B-\beta A)^{-1}B\mu\\
\Sigma_W=&(B-\beta A)^{-1}B \Sigma B(B-\beta A)^{-1}+\frac{\alpha\beta}{2}(B-\beta A)^{-1}
\end{align}
%if igoring the last term (i.e. $\theta=\infty$), the change of measure is equivalent to a coordinate transform by the matrix $B(B-A)^{-1}$. The only requirement is that $B-A$ has to be invertible and positive-definite.

An interesting observation on Eq.~\ref{eq:wvarresult} is that the worst-case measure can be generated by a measure-preserving linear map. In fact, for any vector $v$ of asset returns, the linear map $g$ gives
\begin{samepage}
\begin{align}\label{eq:gv}
g(v)=&(B-\beta A)^{-1}Bv\nonumber\\
=&(I-\beta B^{-1}A)^{-1}v
\end{align}
\end{samepage}
We write down the probability density function for the reference measure by
\begin{align}
f(v)\propto\exp\left(-\frac{1}{2}\left(v-\mu\right)^T\Sigma^{-1}\left(v-\mu\right)\right)
\end{align}
The measure given by the measure-preserving map $g$ has a probability density function that is proportional to $f(g^{-1}(v))$, 
\begin{align}
&f(g^{-1}(v))\nonumber\\
\propto&\exp\left(-\frac{1}{2}\left((I-\beta B^{-1}A)v-\mu\right)^T\Sigma^{-1}\left((I-\beta B^{-1}A)v-\mu\right)\right)\nonumber\\
=&\exp\left(-\frac{1}{2}\left(v-(I-\beta B^{-1}A)^{-1}\mu\right)^T(I-\beta B^{-1}A)\Sigma^{-1}(I-\beta B^{-1}A)\left((I-\beta B^{-1}A)v-\mu\right)\right)\nonumber\\
=&\exp\left(-\frac{1}{2}\left(v-\tilde{\mu}\right)^T\tilde{\Sigma}^{-1}\left(v-\tilde\mu\right)\right)
\end{align}
where
\begin{align}
\tilde{\mu}:=&\left(I-\beta B^{-1}A\right)^{-1}\mu\\
\tilde{\Sigma}:=&\left(I-\beta B^{-1}A\right)^{-1}\Sigma\left(I-\beta B^{-1}A\right)^{-1}
\end{align}
that are precisely the mean and covariance matrix given in Eq.~\ref{eq:wvarresult} (with $\alpha=0$). As a result, we generate the worst-case measure by applying the measure-preserving map $g$.
%Therefore, $g$ is measure-preserving, and if $v$ is a random vector under the reference measure then $g(v)=(I-\beta B^{-1}A)^{-1}v$ follows the distribution function $f_W$.

\subsection{E. The support of a multivariate normal distribution}\label{sec:topo}

In this section, we discuss the support of the reference measure $P$ assuming the asset returns follow a multivariate normal distribution. In addition, we want see how it is altered by different approaches to model risk measurement. Clearly, approaches based on $f$-divergence cannot alter the support as they only account for measures that are equivalent to the nominal one. But this conclusion does not tell us explicitly what the support is. In the following work we aim to find the linear subspace that supports the measure.

Formally speaking, returns of the $n$ assets form a $n$-dimensional vector that lives in a $n$-dimensional topological vector space $\mathcal{V}$. If the asset returns follow a multivariate normal distribution with a non-singular covariance matrix, then the support is the entire space $\mathcal{V}$. However, if the covariance matrix is singular, the support can only be part of $\mathcal{V}$. We will find this support and show that it is a $m$-dimensional linear subspace, where $m$ is the rank of the covariance matrix.

The reference model of asset returns defines a probability space $(\mathcal{V},\mathcal{F},P)$, where $\mathcal{F}$ is the Borel $\sigma$-algebra on $\mathcal{V}$. 
Since $\mathcal{V}$ is a vector space, we may consider its dual space $\mathcal{V}^*$, i.e. the space of linear maps $a: \mathcal{V}\to\mathbb{R}$. Any element of the dual space is regarded as a vector of portfolio weights. To see this, suppose the asset returns are $v=(v_1,v_2,\cdots,v_n)\in\mathcal{V}$, and the portfolio weights are $a=(a_1,a_2,\cdots,a_n)\in\mathcal{V}^*$. The pairing of $a$ and $v$ results in a real number, which is exactly the portfolio return:
\begin{align}
a(v)=\sum_{j=1}^n a_{j}v_j%,~\forall~a\in\mathcal{V}^*,v\in\mathcal{V}
\end{align}

%The reference measure is multivariate normal. 
If we treat the asset returns $v_i$ as random variables, we may calculate of portfolio variance on a given vector of weights $a\in\mathcal{V}$ by $\mathrm{Var}(a(v))=a^T\Sigma a$, where $\Sigma$ is the covariance matrix of the asset returns. For convenience, we use the same symbol $v$ for both the vector of random variables (random vector) and its realization (i.e. a specific element in $\mathcal{V}$). 
Now take the positive semi-definite matrix $\Sigma$ as a linear map $\Sigma:\mathcal{V}^*\to\mathcal{V}$:
\begin{align}
\Sigma(a)=\Sigma a\in\mathcal{V},~\forall a\in\mathcal{V}^*
\end{align}
The portfolio variance is formed by applying the linear map $a:\mathcal{V}\to\mathbb{R}$ to $\Sigma(a)\in\mathcal{V}$: $\mathrm{Var}(a(v))=a(\Sigma(a))$.
If the square matrix $\Sigma$ is singular, then its kernel $\mathrm{ker}\Sigma$ is not trivial (i.e. contains elements other than the zero vector). $\mathcal{V}^*$ can therefore be decomposed into two subspaces: 
\begin{align}\label{eq:decomp}
\mathcal{V}^*=\mathrm{ker}\Sigma\oplus \mathrm{ker}\Sigma^{\perp}
\end{align}
Suppose $\mathrm{ker}\Sigma^{\perp}$ has dimension $n$. $\mathrm{ker}\Sigma$ has dimension $m-n$ for the dimensions of subspaces sum up to the dimension of $\mathcal{V}^*$. We may switch to a new orthonormal basis $\{e^*_1,e^*_2,\cdots,e^*_m,k^*_1,k^*_2,\cdots,k^*_{m-n}\}$ in consistency with the decomposition Eq.~\ref{eq:decomp}, in the sense that $e^*_1, e^*_2,\cdots,e^*_m$ span $\mathrm{ker}\Sigma^{\perp}$ and $k^*_1,k^*_2,\cdots,k^*_{m-n}$ span $\mathrm{ker}\Sigma$. Now get back the original space of asset returns $\mathcal{V}$, we may select a new basis\\
$\{e_1,e_2,\cdots,e_m,k_1,k_2,\cdots,k_{m-n}\}$, dual to $\{e^*_1,e^*_2,\cdots,e^*_m,k^*_1,k^*_2,\cdots,k^*_{m-n}\}$, i.e.
\begin{align*}
e^*_i(e_j)=&\delta_{i-j}\\
k^*_i(k_j)=&\delta_{i-j}\\
e^*_i(k_j)=&0\\
k^*_i(e_j)=&0
\end{align*}
Any $v\in\mathcal{V}$ can be expressed by
\begin{align}\label{eq:expansion}
v=\sum_{i=1}^m u_ie_i+\sum_{i=1}^{m-n}w_ik_i
\end{align}
Suppose $\mathcal{U}$ denotes the linear subspace spanned by $e_1,e_2,\cdots,e_m$. $\mathcal{U}$ is in fact the dual space of $\mathrm{ker}\Sigma^{\perp}$. We will show that the support of the reference measure $P$ is indeed the linear subspace $\mathcal{U}$ shifted by the vector of average asset returns $\mu$:
  \\

Theorem \emph{Given a finite-dimensional topological vector space $\mathcal{V}$ and its Borel $\alpha$-algebra $\mathcal{F}$, the support of a measure $P$ on $(\mathcal{V},\mathcal{F})$ is $\{v\in\mathcal{V}:v-\mu\in\mathcal{U}\}$ if $P$ provides a multivariate distribution $\mathcal{N}(\mu,\Sigma)$.}\\
\\
\emph{Proof}
For every $v\in\mathrm{ker}\Sigma$, consider the variance of $a(v)$ ($v$ is a random vector here):
\begin{align}
\mathrm{Var}(a(v))=a^T\Sigma a=0
\end{align}
The zero variance implies that $a$ carries the measure $P$ on $\mathcal{V}$ to a Dirac measure ${P_a}$ on $\mathbb{R}$
\begin{align}
{P_a}(A)=P(a^{-1}(A)),~\forall A\in\{A\subseteq\mathbb{R}:a^{-1}(A)\in\mathcal{F}\}
\end{align}
Suppose $supp(P_a)=\{s_a\}$ where $s_a\in\mathbb{R}$. We can show that $supp(P)$ should only include elements in $\mathcal{V}$ that is projected to $s_a$. More formally, with the projection map $\mathcal{P}:\mathcal{V}\to\mathrm{ker}\Sigma$, we have
\begin{align}\label{eq:setemp}
\{v\in\mathcal{V}:\exists a\in\mathrm{ker}\Sigma, a(v)\neq s_a\} \cap supp(P)=\emptyset
\end{align}
In fact, for a given $v\in\mathcal{V}$, suppose there exists $a\in\mathrm{ker}\Sigma$ such that $a(v)\neq s_0$. $a(v)$ is not in the $supp(\tilde{P})$, suggesting the existence of an open neighborhood $N_{a(v)}\subseteq\mathbb{R}$ such that ${P_a}(N_{a(v)})=0$. Since the linear map $a$ is continuous, $a^{-1}(N_{a(v)})$ is an open neighborhood of $v$ and
\begin{align}
P(a^{-1}(N_{a(v)}))={P_a}(N_{a(v)})=0
\end{align}
As a result, $v\not\in supp(P)$ which proves Eq.~\ref{eq:setemp}.

Now we consider the set $S:=\{v\in\mathcal{V}:a(v)= s_a, \forall a\in\mathrm{ker}\Sigma\}$. For a given $v_s\in S$, every $v\in S$ satisfies
\begin{align}
a(v-v_s)= a(v)-a(v_s)=0, ~\forall a\in\mathrm{ker}\Sigma
\end{align}
suggesting that $v-v_s\in\mathcal{U}$. Therefore $S=\{v\in\mathcal{V}:v-v_s\in\mathcal{U}\}$. 
Regard $\mathcal{U}$ as a topological linear subspace of $\mathcal{V}$ equipped with the relative topology. $\tilde{\mathcal{F}}$ is the Borel $\sigma$-algebra on $\mathcal{U}$. We may define a new probability space $(\mathcal{U},\tilde{\mathcal{F}},\tilde{P})$ by
\begin{align}
\tilde{P}(A\cap\mathcal{U})=&P(A),~~\forall A\in\mathcal{F}
\end{align}
One can verify that this probability space is well defined. Now we would like to show that $supp(\tilde{P})=\mathcal{U}$. In fact, assuming this is true, then for arbitrary $v\in S$ every open neigborhood $N(v)$ has positive measure:
\begin{align}
P(N(v))=\tilde{P}(N(v)\cap\mathcal{U})>0
\end{align}
This immediately leads to the result $supp(P)=S$. In particular, from the property of the multivariate normal distribution, $supp(P)$ includes the vector $\mu$ of average asset returns. This means that $\mu\in S$, and thus the support of $P$ can be written as $supp(P)=S=\{v\in\mathcal{V}:v-\mu\in\mathcal{U}\}$. 

Now we only need to show that $supp(\tilde{P})=\mathcal{U}$. 
\begin{comment}
the projection map $\mathcal{P}:\mathcal{V}\to \mathcal{U}$ where $\mathcal{U}\subseteq\mathcal{V}$ denote the linear subspace spanned by $\{e_1,e_2,\cdots,e_m\}$. $\mathcal{P}$ sends $(u_1,u_2,\cdots,u_m,w_1,w_2,\cdots,w_{m-n})$ to $(u_1,u_2,\cdots,u_m)$. 
\end{Huge}
The measure induced by the projection is given by
\begin{align}
\tilde{P}(A)=P(\mathcal{P}^{-1}(A))~~\forall A\in\{A\subseteq\mathcal{U}:\mathcal{P}^{-1}(A)\in\mathcal{F}\}
\end{align}
so that $\mathcal{P}$ is a measure-preserving map. 
\end{comment}
Consider the projection map $\mathcal{P}:\mathcal{V}\to\mathcal{U}$ that sends $v=(u_1,u_2,\cdots,u_m,w_1,w_2,\cdots,w_{m-n})$ to $u=(u_1,u_2,\cdots,u_m)$. The projection results in the marginal distribution w.r.t $u_1,u_2,\cdots,u_m$. This marginal distribution characterises a measure $P'$ on the subspace $\mathcal{U}$:
\begin{align}
P'(A)=P(\mathcal{P}^{-1}(A)),~\forall A\in\{A\subseteq\mathcal{U}:\mathcal{P}^{-1}(A)\in\mathcal{F}\}
\end{align}
For any $A\in\mathcal{F}$,  
\begin{align}
P'(A\cap\mathcal{U})=&P(\mathcal{P}^{-1}(A\cap\mathcal{U}))\nonumber\\
=&\tilde{P}(\mathcal{P}^{-1}(A\cap\mathcal{U})\cap\mathcal{U})\nonumber\\
=&\tilde{P}(A\cap\mathcal{U})%, \forall a\in\mathrm{ker}\Sigma
\end{align} 
Therefore, the two measures $\tilde{P}$ and $P'$ coincide, and we only need to prove  $supp(P')=\mathcal{U}$. The marginal distribution from projection $\mathcal{P}$ is apparently multivariate normal (every linear combination of the elements in $u$ is also a linear combination of the elements in $v\in\mathcal{P}^{-1}(u)$ thus normally distributed). 
%Since the random vector $v=(u_1,u_2,\cdots,u_m,w_1,w_2,\cdots,w_{m-n})$ is multivariate normal, every linear combination of its elements, $c_1u_1+c_2u_2+\cdots+c_mu_m+d_1w_1+d_2w_2+\cdots+d_{m-n}w_{m-n}$, is normally distributed. In particular, taking $d_1=d_2=\cdots=d_{m-n}=0$, we conclude that every linear combination of $u_1,u_2,\cdots,u_m$ is normally distributed. As a result, the truncated random vector $u=(u_1,u_2,\cdots,u_m)$ is also multivariate normal.

The covariance matrix $\tilde{\Sigma}$ of the truncated random vector $u$ is invertible. 
%To see this point, consider $\tilde{\Sigma}$ as a linear map $\tilde{\Sigma}:\mathrm{ker}\Sigma^{\perp}\to\mathcal{U}$. 
In fact, because $\Sigma(a)$ is not a zero vector for every non-zero $a\in\mathrm{ker}\Sigma^\perp$, the linear map between two $m$-dimensional vector spaces $\Sigma|_{\mathrm{ker}\Sigma^\perp}:\mathrm{ker}\Sigma^\perp\to\Sigma(\mathrm{ker}\Sigma^\perp)$ is invertible. Represented by a $m\times m$ matrix, $\Sigma|_{\mathrm{ker}\Sigma^\perp}$ has only non-zero eigenvalues. Since it is also positive semi-definite (for $\mathrm{Var}(a(v))=a(\Sigma(a))\geq0,~\forall a\in\mathrm{ker}\Sigma^\perp\subseteq\mathcal{V}^*$), it must be positive definite. We conclude that for every non-zero $a\in\mathrm{ker}\Sigma^\perp$, $\mathrm{Var}(a(v))=a(\Sigma|_{\mathrm{ker}\Sigma^\perp} (a))>0$. If we expand $a(v)$ component-wise according to Eq.~\ref{eq:expansion},
\begin{align}
a(v)=&\sum_{i=1}^m u_ia(e_i)+\sum_{i=1}^{m-n}w_ia(k_i)\nonumber\\
=&a\left(\sum_{i=1}^m u_ie_i\right)=a(u)
\end{align}
Therefore $a^T\tilde{\Sigma}a=\mathrm{Var}(a(u))=\mathrm{Var}(a(v))>0$ for every nonzero $a\in\mathrm{ker}\Sigma^\perp$. %  the truncated random vector $u=(u_1,u_2,\cdots,u_m)$. 
As a result, $\tilde{\Sigma}a$ is positive-definite and thus invertible. Under the measure $P'$, the random vector $u$ follows a multivariate normal distribution with a non-singular covariance matrix. It is supported by the entire subspace $\mathcal{U}$, i.e. $supp(P')=\mathcal{U}$. $\square$

\begin{comment}
the support of the measure is a subspace of $\mathcal{V}$. We may denote the subspace by $\mathcal{U}$ and decompose $\mathcal{V}=\mathcal{U}\oplus\mathcal{U}^{\perp}$. Each asset return can be projected to the two subspaces and decompose into $X_i=U_i+U_i^{\perp}$. Assuming $w\in\mathcal{U}^*$ and $w'\in\mathcal{U}^*$, 

Take the covariance matrix as an operator $\Sigma:\mathcal{V}^*\to\mathcal{V}^*$.  $\mathcal{U}^{\perp}*$ is the kernel of $\Sigma$. In fact, the portfolio variance is zero (under the reference model) if we hold positions according to a covector $a^{\perp}\in\mathcal{U}^{\perp}*$, i.e.
\begin{align}
\langle a|\Sigma a\rangle=0,~\forall~ a^{\perp}\in\mathcal{U}^{\perp}*
\end{align}

The support of the measure is $\Sigma^{1/2}v$. 

However, the covector
\begin{align}
Var(w'(V))=
\langle w'|\Sigma w'\rangle=0,~V\in\mathcal{U}
\end{align}
meaning that $w'$ maps all vectors in $\mathcal{U}$ to a constant. This is parallel to kernel of the inverse map.

\end{comment}

For a multivariate distribution $\mathcal{N}(\mu,\Sigma)$, 
the support $supp(P)=\{v\in\mathcal{V}:v-\mu\in\mathcal{U}\}$ only depends on the vector $\mu$ and the kernel of $\Sigma$. It is clear that under the Kullback-Leibler divergence the worst-case measure shares the same support. 
In fact, the worst-case distribution is $\mathcal{N}(\mu_{KL},\Sigma_{KL})$ where $\mu_{KL}$ and $\Sigma_{KL}$ are given in Eq.~\ref{eq:varkl}. Assuming $\theta$ is sufficiently small so that $I-2\theta \Sigma A$ is invertible, $\Sigma_{KL}a=0$ if and only if $\Sigma_{KL}=0$ for every $a\in\mathcal{V}^*$. Therefore, $\Sigma_{KL}$ and $\Sigma$ share the same kernel and therefore the same subspace $\mathcal{U}\subseteq\mathcal{V}$. In addition, $\mu_{KL}-\mu\in\mathcal{U}$ because for every $a\in\mathrm{ker}\Sigma$ we have
\begin{align}
a(\mu_{KL}-\mu)=&a\left(2\theta \Sigma A(I-2\theta \Sigma A)^{-1}\mu\right)\nonumber\\
=&2\theta(\Sigma a)^TA(I-2\theta \Sigma A)^{-1}\mu\nonumber\\
=&0
\end{align}
As a result, the support of the worst-case measure is $\{v\in\mathcal{V}:v-\mu_{KL}\in\mathcal{U}\}=\{v\in\mathcal{V}:v-\mu\in\mathcal{U}\}$, same as the support of the reference measure.

On the other hand, the worst-case measured resulted from the Wasserstein approach can have different support. According to Eq.~\ref{eq:varw}, the worst-case covariance matrix $\Sigma_W$ has a different kernel in general. In addition, $\mu_{W}-\mu=\beta A(B-\beta A)^{-1}\mu$ is not dependent on $\Sigma$, thus not linked to the subspace $\mathcal{U}$. Setting $\alpha=0$ in Eq.~\ref{eq:varw} provides a particularly interesting case, where the worst-case measure is given by a measure-preserving linear map $g:\mathcal{V}\to\mathcal{V}$ given by Eq.~\ref{eq:gv}. 
As a result, the support of the worst-case measure can be obtained using the same map, i.e.
\begin{align}
&\left\{v\in\mathcal{V}:g^{-1}(v)-\mu\in\mathcal{U}\right\}\nonumber\\=&
\left\{v\in\mathcal{V}:(I-\beta B^{-1}A)v-\mu\in\mathcal{U}\right\}\nonumber\\
=&
\left\{v\in\mathcal{V}:v-(I-\beta B^{-1}A)^{-1}\mu\in\{(I-\beta B^{-1}A)^{-1}u:u\in\mathcal{U}\}\right\}\nonumber\\
=&
\left\{v\in\mathcal{V}:v-\mu_W\in\mathcal{U}_W\right\}
\end{align}
$\mathcal{U}_W:=\{(I-\beta B^{-1}A)^{-1}u:u\in\mathcal{U}\}\subseteq\mathcal{V}$ is the linear subspace (perpendicular to $\mathrm{ker}\Sigma_W$) that corresponds to the worst-case scenario under the Wasserstein approach.

\subsection{F. Verification of the Wasserstein approach}
Sec.~\ref{sec:topo} shows that under the Wasserstein approach the worst-case measure does alter the support. Now the question is whether the approach searches over all alternative measures. Unlike $f$-divergence that is only capable of measuring distance between equivalent measures, the Wasserstein metric provides a finite distance between non-equivalent measures as well. Therefore the Wasserstein approach should be able to find out the worst-case measure from all equivalent and non-equivalent measures. In this section, we will verify it for the example of portfolio variance. In particular, we will find out a worst-case linear map $g^*:\mathcal{V}\to\mathcal{V}$ by searching over the entire space of linear maps. We will verify that Eq.~\ref{eq:varkl} (with $\alpha=0$) can be given by the worst-case linear map.

Theorem \emph{Given a probability space $(\mathcal{V},\mathcal{F},P)$ where $\mathcal{V}$ is a finite-dimensional vector space and $P$ provides a multivariate distribution $\mathcal{N}(\mu,\Sigma)$, there exists a worst-case linear map $g^*:\mathcal{V}\to\mathcal{V}$ in the sense of Eq.~\ref{eq:ystar}, i.e.
\begin{align}\label{eq:gstarx}
g^*(x)=\arg\max_{y\in\mathcal{V}}\left[y^TAy-\frac{(x-y)^TB(x-y)}{\beta}\right]
\end{align}
}
for every non-zero $x\in\mathcal{V}$, as long as $B-\beta A$ is positive definite.\\
\emph{Proof}
Given a non-zero $x\in\mathcal{V}$, every non-zero $y\in\mathcal{V}$ can be expressed by $y=g(x)$ where $g$ is some linear map (not unique) $g:\mathcal{V}\to\mathcal{V}$. The problem Eq.~\ref{eq:gstarx} is therefore equivalent to
\begin{align}\label{eq:gxmax}
g^*(x)=\arg\max_{g\in\mathfrak{L}(\mathcal{V},\mathcal{V})}\left[g(x)^TAg(x)-\frac{\left(x-g(x)\right)^TB\left(x-g(x)\right)}{\beta}\right](x)
\end{align}
where $\mathfrak{L}(\mathcal{V},\mathcal{V})$ is the space of all linear maps from $\mathcal{V}$ to $\mathcal{V}$.
Choosing a orthonormal basis for $\mathcal{V}$ allows us to represent $g$ by a square matrix, and the linear map $g(x)$ by matrix multiplication $gx$. The expression inside the square bracket in Eq.~\ref{eq:gxmax} is then transformed into 
\begin{align}\label{eq:calcgx}
&(gx)^TAgx-\frac{\left(x-gx\right)^TB\left(x-gx\right)}{\beta}\nonumber\\
=&-\frac{1}{\beta}x^T\left(g^T-B(B-\beta A)^{-1}\right)(B-\beta A)\left(g-(B-\beta A)^{-1}B\right)x\\
&-\frac{1}{\beta}x^T \left(B-B(B-\beta A)^{-1}B\right)x\nonumber
\end{align}
Since $B-\beta A$ is positive-definite, the first term in Eq.~\ref{eq:calcgx} is either zero or negative. It reaches zero (and hence Eq.~\ref{eq:calcgx} reaches its maximum value) if and only if
\begin{align}
\left(g-(B-\beta A)^{-1}B\right)x=0
\end{align}
or equivalently
\begin{align}
g(x)=(B-\beta A)^{-1}Bx
\end{align}
This allows to rewrite Eq.~\ref{eq:gxmax} by
\begin{align}\label{eq:gstarex}
g^*(x)=(B-\beta A)^{-1}Bx
\end{align}
The linear map $g^*$ given by the square matrix $(B-\beta A)^{-1}B$ satisfies Eq.~\ref{eq:gstarex} and thus solves Eq.~\ref{eq:gxmax} for every non-zero $x\in\mathcal{V}$.
$\square$

It is noted that in the problem of portfolio variance risk, both square matrices $A$ and $B$ are symmetric and positive definite. Therefore if the positive multiplier $\beta$ is sufficiently small, $B-\beta A$ is also positive definite satisfying the condition assumed in the theorem above. 
Now the worst-case linear map $g^*$ transforms the vector of asset returns from $\mu$ to $(B-\beta A)^{-1}B\mu$, and the covariance matrix from $\Sigma$ to $(B-\beta A)^{-1}B\Sigma B(B-\beta A)^{-1}$, same as the expressions given in Eq.~\ref{eq:varkl} (with $\alpha=0$). This verifies that the Wasserstein approach indeed searches over  the entire space $\mathfrak{L}(\mathcal{V},\mathcal{V})$ of linear maps. It results in a measure that corresponds to the worst-case linear map $g^*$.

\begin{comment}
It in fact searches over all $m'$-dimensional subspace of $\mathcal{V}$ ($m'\leq m$). Denote the support of the reference measure by $\mathcal{U}\sim\mathbb{R}^m$, spanned by a basis $\{v_i\}_{1\leq i\leq m}$. Alternative subspace of the same dimension is spanned by $\{u_i\}_{1\leq i\leq m}$. One can find a linear map (not unique) $T:\mathcal{V}\to\mathcal{V}$ that maps $v_i$ to $u_i$ ($i=1,2,\cdots,m$). 
Therefore, for any $m$-dimensional subspace $\mathcal{U}'$, there exists a linear map $T$ that maps $\mathcal{U}$ onto $\mathcal{U}'$. This corresponds to the fact that the Wasserstein approach can alter the support. Among all these $m$-dimensional subspaces, provides the one that corresponds to the worst-case scenario.
\end{comment}

\subsection{G. Robust MVO Portfolio (Kullback-Leibler divergence)}
According to Eq.~\ref{eq:problemrobust}, we consider the problem
\begin{align}
\max_{Q\in\mathscr{M}} E^Q\left((X-\mu-k)^Taa^T(X-\mu-k)\right)
\end{align}
Since $X-\mu-k\sim\mathcal{N}(-k,\Sigma)$, under the Kullback-Leibler divergence the covariance matrix and the mean of the worst-case measure are given according to Eq.~\ref{eq:varkl} (remember $a^Tk=\lambda/2$): 
\begin{align}\label{eq:klworstcase}
\Sigma_{KL}=&(I-2\theta\Sigma A)^{-1}\Sigma\\
\mu_{KL}=&(\mu+k)-\left(I-2\theta\Sigma A\right)^{-1}k\nonumber\\
=&\mu-\lambda\theta(I-2\theta\Sigma A)^{-1}\Sigma a\nonumber
\end{align}
Using the Wassertein approach, however, the worst-case measure has different covariance matrix and mean (Eq.~\ref{eq:varw}):
\begin{align}\label{eq:wworstcase}
\Sigma_{W}=&(I-\beta B^{-1} A)^{-1}\Sigma(I-\beta AB^{-1} )^{-1}\\
\mu_{W}=&(\mu+k)-(I-\beta B^{-1} A)^{-1}k\nonumber\\
=&\mu-\frac{\lambda}{2}\beta(I-\beta B^{-1} A)^{-1}B^{-1} a\nonumber
\end{align}
We may then formulate the optimal asset allocation $a^*$ under the worst-case measure. According to Eq.~\ref{eq:mina}, the problem is formulated in the following form under the Kullback-Leibler divergence.
\begin{align}
\min_a \,&a^T\Sigma_{KL} a-\lambda a^T\mu_{KL}\nonumber\\
=&a^T(I-2\theta\Sigma A)^{-1}\Sigma a-\lambda a^T(\mu-\lambda\theta(I-2\theta\Sigma A)^{-1}\Sigma a)\nonumber\\
=&a^T\Sigma a-\lambda a^T\mu+\theta a^T\Sigma a\left(\lambda^2+2 a^T\Sigma a\right)+O(\theta^2)
\end{align}
Note that in the last equality we apply the Taylor expansion $(I-2\theta\Sigma A)^{-1}=I+2\theta\Sigma A+4\theta^2\Sigma A\Sigma A+\cdots=I+2\theta\Sigma A+O(\theta^2)$. 
To find out a closed-form solution, we need to ignore the higher order terms $O(\theta^2)$. Then the stationary condition of the minimisation problem is given by a non-linear equation:
\begin{align}\label{eq:klequation}
2\Sigma a-\lambda \mu+2\theta \left(\lambda^2+4a^T\Sigma a\right)\Sigma a
=0
\end{align}

Notice that
\begin{align}
a^*:=\frac{\lambda}{2}\Sigma^{-1}\mu
\end{align}
is the MVO portfolio weight under the reference measure. For the robust MVO portfolio, we may consider its first-order deviation from $a^*$. To do that, we substitute $a=a^*+\theta b$ into Eq.~\ref{eq:klequation} allowing us to cancel the term $\lambda\mu$. 
\begin{align}
2\theta\Sigma b+\theta\lambda\left(\lambda^2+\lambda^2\mu^T\Sigma^{-1}\mu\right)\mu+O(\theta^2)=0
\end{align}
By matching the first-order term w.r.t $\theta$, we find the expression for $b$:
\begin{align}
b=-\frac{\lambda^3}{2}\left(1+\mu^T\Sigma^{-1}\mu\right)\Sigma^{-1}\mu
\end{align}
Therefore the optimal MVO portfolio under the worst-case scenario is
\begin{align}\label{eq:aKL}
a^*_{KL}=&\left(\frac{\lambda}{2}-\frac{\theta\lambda^3}{2}\left(1+\mu^T\Sigma^{-1}\mu\right)\right)\Sigma^{-1}\mu\nonumber\\
=&c a^*
\end{align}
where the coefficient $c$ is defined by
\begin{align}
c:=1-\theta\lambda^2\left(1+\mu^T\Sigma^{-1}\mu\right)
\end{align}

The robust MVO portfolio, as a vector $a^*_{KL}$, is parallel to the normal MVO portfolio $a^*$. As a result, the robust MVO portfolio does not change the relative weights of component assets. In fact, all the weights are reduced by the same proportion ($c<1$) to account for model risk. This is, however, inappropriately account for the correlation risk. For example, two highly-correlated assets have extremely high weights in the nominal MVO portfolio. Because of the correlation risk, we would expect the robust MVO portfolio to assign them lower weights relative to other assets.

The Sharpe ratio of the robust MVO portfolio obviously equals the Sharpe ratio under the reference measure, denoted by $S$ ($S=\sqrt{\mu^T\Sigma^{-1}\mu}$). Sometimes we may be interested in the Sharpe ratio under the worst-case measure.  
This requires us to examine the mean and variance of the robust MVO portfolio given by Eq.~\ref{eq:aKL}. Assuming that we are under the worst-case scenario given by Eq.~\ref{eq:klworstcase}, the portfolio mean and variance can be obtained by substituting $a^*_{KL}=c\lambda\Sigma^{-1}\mu/2$:
\begin{align}\label{eq:meanvar}
\mu_{KL}^Ta^*_{KL}=&\left(\mu-\lambda\theta\left(I-2\theta\Sigma a^*_{KL}a^{*T}_{KL}\right)^{-1}\Sigma a^*_{KL}\right)^Ta^*_{KL}\\
=&
\frac{\lambda c}{2}\mu^T\Sigma^{-1}\mu-\theta\frac{\lambda^3c^2}{4} \mu^T\Sigma^{-1}\mu +O(\theta^2)\\
a^{*T}_{KL}\Sigma_{KL}a^*_{KL}=&a^{*T}_{KL}\left(I-2\theta\Sigma a^*_{KL}a^{*T}_{KL}\right)^{-1}\Sigma a^*_{KL}\\
=&
\frac{\lambda^2c^2}{4}\mu^T\Sigma^{-1}\mu+\theta \frac{\lambda^4c^4}{8}(\mu^T\Sigma^{-1}\mu)^2+O(\theta^2)
\end{align}
By using the portfolio mean and variance given in Eq.~\ref{eq:meanvar}, we may calculate the Sharpe ratio of the robust MVO portfolio (under the worst-case scenario):
\begin{align}
S_{KL}=&\frac{\mu_{KL}^Ta^*_{KL}}{\sqrt{a^{*T}_{KL}\Sigma_{KL}a^*_{KL}}}\nonumber\\
=&
\frac{2-\theta \lambda^2c+O(\theta^2)}{2+\frac{\theta}{2}\lambda^2c^2\mu^T\Sigma^{-1}\mu+O(\theta^2)}\sqrt{\mu^T\Sigma^{-1}\mu}\nonumber\\
=&\left(1-\frac{\theta}{4}\lambda^2c\left(cS^2+2\right)
+O(\theta^2)\right)S
\end{align}
We can see that the robust Sharpe ratio (defined as the Sharpe ratio of the robust MVO portfolio under the worst-case model) is a function of the nominal Sharpe ratio $S$. The MVO portfolio corresponds to $c=1$, suffering from more reduction in Sharpe ratio than the robust MVO portfolio ($c<1$) under the worst-case measure. This simple relation, however, no longer holds for the Wasserstein approach.

\subsection{H. Robust MVO Portfolio (Wasserstein approach)}
In this section, we will switch to the Wasserstein approach to model risk measurement. We will derive the robust MVO portfolio with the Wasserstein approach. Using Eq.~\ref{eq:wworstcase}, we may formulate the robust portfolio optimisation problem in the following form:
\begin{align}
\min_a\,&a^T\Sigma_Wa-\lambda a^T\mu_W\nonumber\\
=&a^T(I-\beta B^{-1}A)^{-1}\Sigma(I-\beta AB^{-1})^{-1}a-\lambda a^T\left(\mu-\frac{\lambda}{2}\beta(I-\beta B^{-1} A)^{-1}B^{-1} a\right)\nonumber\\
=&a^T\Sigma a-\lambda a^T\mu+\beta \left(2a^TB^{-1}aa^T\Sigma a%+a^T\Sigma a a^T\right]
+\frac{\lambda^2}{2} a^TB^{-1}a\right)+O(\beta^2)
\end{align}
Ignoring the higher order terms, the minimisation problem is solved using
\begin{align}\label{eq:wassereq}
2\Sigma a-\lambda \mu+\beta\left(4a^TB^{-1}a\Sigma a+(4a^T \Sigma a+\lambda^2)B^{-1}a\right)=0
\end{align}
Substituting $a=a^*+\beta b$ into Eq.~\ref{eq:wassereq}, we find the expression for the perturbation $b$ by matching the first-order terms of $\beta$:
\begin{align}
b=-\frac{\lambda^3}{4}\left(\mu^T\Sigma^{-1}B^{-1}\Sigma^{-1}\mu+\left(1+\mu^T\Sigma^{-1}\mu\right)\Sigma^{-1}B^{-1}\right)\Sigma^{-1}\mu
\end{align}
Therefore, the Wasserstein approach results in a robust MVO portfolio with weights 
\begin{align}\label{eq:wweights}
a^*_W=&\left(\frac{\lambda}{2}-\frac{\beta\lambda^3}{4}\mu^T\Sigma^{-1}B^{-1}\Sigma^{-1}\mu-\frac{\beta\lambda^3}{4}\left(1+\mu^T\Sigma^{-1}\mu\right)\Sigma^{-1}B^{-1}\right)\Sigma^{-1}\mu\nonumber\\
=&ca^*-Da^*
\end{align}
where $c$ is a coefficient while $C$ is a square matrix defined by
\begin{align}
c:=&1-\frac{\beta\lambda^2}{2}\mu^T\Sigma^{-1}B^{-1}\Sigma^{-1}\mu\nonumber\\
D:=&\frac{\beta\lambda^2}{2}\left(1+\mu^T\Sigma^{-1}\mu\right)\Sigma^{-1}B^{-1}:=d\Sigma^{-1}B^{-1}
\end{align}
$c$ serves just as the coefficient under the Kullback-Leibler divergence, reducing the portfolio weights by the same fraction. $D$ is a matrix that serves to linearly transform the normal MVO portfolio weights.

Eq.~\ref{eq:wweights} correctly accounts for the correlation risk. When two assets are highly correlated, $\Sigma$ is close to be singular. This results in extremely large weights under the normal MVO portfolio. Eq.~\ref{eq:wweights}, on the other hand, not only scales the weights down simultaneously by the coefficient $c$, but also reduces  the relative weights of the highly-correlated assets by the linear map $D$. To see how the linear map $D$ changes the relative weights, we may re-arrange 
Eq.~\ref{eq:wweights} to the following form:
\begin{align}
a_W^*=\frac{\lambda}{2}\left(\Sigma(cI-D)^{-1}\right)^{-1}\mu
\end{align}
Therefore, the robust MVO portfolio has the same weights as a normal MVO portfolio with an effective covariance matrix
\begin{align}
\Sigma^*=\Sigma(cI-d\Sigma^{-1}B^{-1})^{-1}
\end{align}
One can show by induction that $\Sigma v=xv$ ($x$ and $v$ are respectively the eigenvalue and the eigenvector) leads to $\Sigma^n v=xv$ for every integer $n$. This is to say, $x$ is an eigenvalue of $\Sigma$ only if $x^n$ is an eigenvalue of $\Sigma^n$ corresponding to the same eigenvector. As a result, for every eigenvalue $x>0$ of the positive-definite covariance matrix, there exists a corresponding eigenvalue of the effective covariance matrix (here we only consider the special case where $B$ is the identity matrix $I$):
\begin{align}
\Sigma^*v=&\Sigma(cI-d\Sigma^{-1})^{-1}v\nonumber\\
=&\frac{1}{c}\sum_{i=0}^\infty \left(\frac{d}{c}\right)^i\Sigma^{1-i}v\nonumber\\
=&\frac{1}{c}\sum_{i=0}^\infty \left(\frac{d}{c}\right)^ix^{1-i}v\nonumber\\
=&\frac{x}{c-{d}/{x}}v
\end{align}
The corresponding eigenvalue 
\begin{align}\label{eq:adjust}
x^*:=&\frac{x}{c-{d}/{x}}\nonumber\\
=&x+\frac{\beta\lambda^2}{2}\left(1+\mu^T\Sigma^{-1}\mu+\mu^T\Sigma^{-1}B^{-1}\Sigma^{-1}\mu\right)+O(\beta^2)
\end{align}
Any eigenvalue $x$ close to zero is adjusted according to Eq.~\ref{eq:adjust}, resulting in a corresponding eigenvalue $x^*$ that is at least as large as $\beta\lambda^2/2$. This results in an effective matrix $\Sigma^*$ that is less ''singular" than $\Sigma$, and therefore a robust MVO portfolio that accounts for the correlation risk.

\begin{figure}[H]
\begin{center}
\includegraphics[width=.6\textwidth]{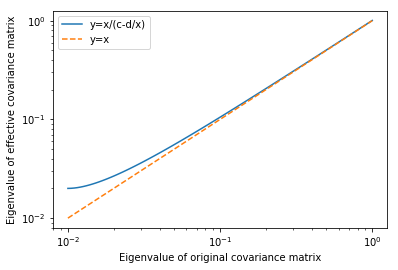}
\renewcommand{\figurename}{Fig}
\caption{Eigenvalue $x^*$ of the effective covariance matrix $\Sigma^*$ increases by a greater amount when the original eigenvalue $x$ gets closer to zero.}
\end{center}
\end{figure}

\begin{comment}
The optimal robust portfolio has the mean and the covariance matrix of
\begin{align}
%\mu^TC\Sigma^{-1}\mu-
\mu^T\Sigma^{-1}C\left[I-\beta (B-\beta A)^{-1}C\Sigma^{-1}\right]\mu\\
\mu^T\Sigma^{-1}C(I-\beta B^{-1} A)^{-1}\Sigma(I-\beta AB^{-1} )^{-1}C\Sigma^{-1}\mu
\end{align}
which provides a Sharpe ratio that is way more robust (on extreme correlations).
\end{comment}

\bibliographystyle{unsrt}
\bibliography{ref}

\end{document}